\documentclass[11pt]{article}

\usepackage[utf8]{inputenc}
\usepackage[T1]{fontenc}
\usepackage{newpxtext}

\usepackage{tcolorbox}
\usepackage{fullpage}
\usepackage{tikz}
\usepackage{algorithmicx}
\usepackage{algorithm}
\usepackage[noend]{algpseudocode}
\usepackage[colorinlistoftodos]{todonotes}
\usepackage{soul}
\usepackage{url}
\definecolor{purple}{cmyk}{.51,.91,0,.34}
\usepackage{microtype}
\usepackage[pdftex, colorlinks,
  linkcolor=blue,
  citecolor=purple,
  filecolor=blue,urlcolor=blue]%
  {hyperref}

\makeatletter
\newcommand{\leqnos}{\tagsleft@true\let\veqno\@@leqno}
\newcommand{\reqnos}{\tagsleft@false\let\veqno\@@eqno}
\makeatother

\makeatletter
\newcommand\newtag[2]{#1\def\@currentlabel{#1}\label{#2}}
\makeatother

\usepackage{amsfonts,amssymb}
\usepackage{mathrsfs}
\usepackage{mathtools}
\usepackage{amsthm}
\newtheorem{theorem}{Theorem}[section]
\newtheorem{lemma}[theorem]{Lemma}

\newtheorem{corollary}[theorem]{Corollary}
\newtheorem{claim}[theorem]{Claim}
\newtheorem{open-problem}[theorem]{Open Problem}

\theoremstyle{definition}
\newtheorem{definition}[theorem]{Definition}
\newtheorem{conjecture}[theorem]{Conjecture}
\newtheorem{example}[theorem]{Example}
\newtheorem{remark}[theorem]{Remark}
\newtheorem{observation}[theorem]{Observation}
\theoremstyle{remark}
\newtheorem{problem}[theorem]{Problem}
\newcommand{\NP}{NP}
\newcommand{\pf}{{\textbf {Proof:} }}
\numberwithin{equation}{section}

\usepackage{lineno}

\title{Digraphs Homomorphism Problems with Maltsev Conditions }

 \author{
       Jeff Kinne \thanks{Indiana State University, IN, USA, jkinne@cs.indstate.edu, supported by NSF 1751765} 
      \and 
  Ashwin Murali \thanks{Indiana State University, IN, USA, amurali1@sycamores.indstate.edu, supported by NSF 1751765}
\and
   Arash Rafiey \thanks{Indiana State University, IN, USA arash.rafiey@indstate.edu and Simon Fraser University, BC, Canada, arashr@sfu.ca, supported by NSF 1751765}}

\date{}

\begin{document} 

\maketitle
\begin{abstract}
We consider a generalization of finding a homomorphism from an input digraph $G$ to a fixed digraph $H$, HOM($H$). 
In this setting, we are given an input digraph $G$ together with a list function from $G$ to $2^H$. The goal is to find a homomorphism from $G$ to $H$ with respect to the lists if one exists. 
 
We show that if the list function is a Maltsev polymorphism then deciding whether $G$ admits a homomorphism to $H$ is polynomial time solvable. In our approach, we only use the existence of the Maltsev polymorphism. Furthermore, we show that 
deciding whether a relational structure $\mathcal{R}$ admits a Maltsev polymorphism is a special case of finding a homormphism from a graph $G$ to a graph $H$ and a list function with a Maltsev polymorphism.  Since the existence of Maltsev is not required in our algorithm, we can decide in polynomial time whether the relational structure $\mathcal{R}$ admits  Maltsev or not. 

We also discuss forbidden obstructions for the instances admitting Maltsev list polymorphism. We have implemented our algorithm and tested on instances arising from linear equations, and other types of instances.



\end{abstract}


\section{Motivation and Background} \label{apsec:background}




\paragraph{CSP Motivation}
The \emph{constraint satisfaction problem (CSP)} involves deciding, given a set
of variables and a set of constraints on the variables, whether or not there is an assignment to
the variables satisfying all of the constraints. This problem can be formulated in terms of homomorphims as follows. Given a pair $(\mathcal{G}, \mathcal{H})$ of \emph{relational structures}, decide whether or not there is
a homomorphism from the first structure to the second structure. 
A common way to restrict this problem is to fix the second structure $\mathcal{H}$, so that each structure $\mathcal{H}$ gives
rise to a problem CSP($\mathcal{H}$). The most effective approach to the study of the CSP($\mathcal{H})$ is the so-called
algebraic approach that associates every $\mathcal{H}$ with its \emph{polymorphisms}. 
Roughly speaking, the presence of nice enough polymorphisms leads directly to polynomial time tractability of CSP($\mathcal{H}$), while their absence leads to hardness.  
Beside decision CSPs, polymorphisms have been used extensively for approximating CSPs, robust satisfiability of CSPs, and testing solutions (in the sense of property testing)~\cite{ChenVY16,DalmauKKMMO17,esa2012,kotyz,RRS}.

An interesting question arising from these studies, in particular the CSP Dichotomy Theorem~\cite{B17,arash-dichotomy,Z17}, is known as the \emph{meta-question}. Given a relational structure $\mathcal{H}$, decide whether or not $\mathcal{H}$ admits a polymorphism from a class--for various classes of polymorphims. For many cases hardness results are known. Semmilattice, majority, Maltsev, near unanimity, and weak near unanimity, are among the popular polymorphisms when it comes to study of CSP. Having one or more of these polymorphisms on relation $\mathcal{H}$, would make the  CSP($\mathcal{H}$) (or variation) instance tractable. Therefore, knowing structural characterization and polynomial time recognition for these polymorphisms would help in designing efficient algorithms for CSP.   

It was shown in \cite{benoit}  that deciding if a relational structure admits any of the following polymorphism is NP-complete;  a semilattice polymorphism, a conservative semilattice polymorphism,  a commutative, associative polymorphism (that is, a commutative semigroup polymorphism). However, when $H$ is a digraph then deciding whether $H$ admits a conservative semmilattice is polynomial time solvable \cite{bi-arc}. Relational structure and digraphs with majority/ near unanimity polymorphism have studied in  \cite{nuf1,benoit,nuf2,soda11,kazda,maroti}.


However, deciding whether a relational structure $R$ admit a Maltsev proven to be more challenging. 
Some very interesting works, using algebraic techniques, have been developed on the complexity of deciding whether an idempotent algebra has Maltsev term \cite{matt-v1,Ryerson,mayr,Siggers}.

One remaining open question is an efficient procedure to recognize whether an input relational structure admits a Maltsev polymorphism (not necessarily conservative). \\

The presence of Maltsev polymorphisms lead to several positive results. As an example, it is now a classic theorem in the area that for any structure $\mathcal{H}$ having a Maltsev polymorphism, the problem CSP($\mathcal{H}$) is polynomial time decidable \cite{BD06}. 

\paragraph{Graph Theory Motivation} 

A {\em homomorphism} of a digraph $G$ to a digraph $H$ is a mapping $g$ of the vertex set of $G$ to the vertex set of $H$
so that for every arc $uv$ of $G$,  $g(u)g(v)$, the image of $uv$, is an arc of $H$. A natural decision problem is whether for given
digraphs $G$ and $H$ there is a homomorphism from $G$ to $H$.  If we view (undirected) graphs as digraphs in which each edge
is replaced by two opposite directed arcs, we may apply the definition to graphs as well. An easy reduction from the
$k$-coloring problem shows that this decision problem is $\NP$-hard: a graph $G$ admits a $3$-coloring if and only if there is
a homomorphism from $G$ to $K_3$, the complete graph on $3$ vertices. As a homomorphism is easily verified if the mapping
is given, the homomorphism problem is contained in $\NP$ and is thus $\NP$-complete.

For a fixed digraph $H$ the problem $HOM(H)$ asks
if a given input digraph $G$ admits a homomorphism to $H$. Note that while the above reduction shows $HOM(K_3)$ is NP-complete,
$HOM(H)$ can be easy (in $P$) for some graphs $H$: for instance if $H$ contains a vertex with a self-loop, then every graph $G$ admits a homomorphism
to $H$. Less trivially, for $H = K_2$ (or more generally, for any bipartite graph $H$), there is a homomorphism from $G$ to $K_2$ if
and only if $G$ is bipartite. A very natural goal is to identify precisely for which digraphs $H$ the problem $HOM(H)$ is polynomial time solvable. 

The {\em list homomorphism problem} for digraph $H$, LHOM($H$) is a generalization of the homomorphism problem. We are given an input digraph $G$ together with the lists, $L$ where for every $x \in V(G)$, $L(x) \subseteq V(H)$, and the goal is to find a homomorphism $G$ to $H$ with respect to the lists; the image of each vertex of $G$ must be in its list. List homomorphism problems are known to have nice dichotomies \cite{FH98,FHH03,FHH07,soda11}. However, in our general setting $H$ itself may not have any of the special polymorphisms, and hence,  we can not used those results in our approach. 

The existence of 
conservative polymorphisms is a hereditary property (if $H$ has a particular kind of conservative 
polymorphism, then so does any induced subgraph of $H$). Thus, these questions present interesting 
problems in graph theory. 
In terms of obstruction for various polymorphism,
there are forbidden obstruction
characterizations for existence of conservative majority \cite{soda11} and conservative Maltsev 
 polymorphisms in digraphs \cite{catarina,soda11}.




\section{Our Results } 

Our main result is the following theorem. 

\begin{theorem}\label{Maltsev-poly}
 Let $G$ and $H$ be two digraphs and let $L : V(G) \rightarrow 2^{H}$. Then  $HOM(H)$ for $G$ with respect to $L$, is polynomial time solvable  (in terms of  both $G,H$) when $G \times_L H^3$ admits a Maltsev polymorphism and without
 knowing the actual value of the Maltsev polymorphism. 
  \label{thm:main}
\end{theorem}

We will show that the algorithm does not require being given the Maltsev list polymorphism; existence is sufficient. We also will see that the algorithm is polynomial regardless of whether $H$ admits a Maltsev polymorphism, and the algorithm is correct as long as there is a Maltsev list polymorphism on  $G \times_L H^3$.

Using Theorem \ref{Maltsev-poly} we prove the following theorem which is another main results of this paper. 

\begin{theorem}\label{maltsev-regontition}
Let $\mathcal{R}$ be a relational structure. Then deciding whether $\mathcal{R}$ admits a Maltsev polymorphism is polynomial time solvable. 
\end{theorem}

Notice that, in the usual setting, it was proved in \cite{BD06} that if a relational structure $\mathcal{R}$ admits a Maltsev polymorphism then CSP($\mathcal{R}$) is polynomial time solvable. We will show examples where digraph $H$ itself does not admit a Maltsev while $G \times_L H^3$ admit a Maltsev polymorphism.

\section{ Preliminaries and Notation}

Let $G$ be a digraph.  We let $V(G)$ and $A(G)$ denote the 
vertices and arcs (or edges) of $G$.  In place of $(u,v) \in A(G)$ we use the shorthand - 
$uv\in A(G)$ or $uv\in G$.  In place of $u\in V(G)$ we use the shorthand - $u\in G$. 

For digraphs $G_1,G_2,\dots,G_k$, let $G_1 \times G_2 \times \dots \times G_k$ be the digraph with vertex set $\{(x_1,x_2,\dots,x_k) \mid x_i \in G_i, 1 \le i \le k\}$ and arc set $\{(x_1,x_2,\dots,x_k)(x'_1,x'_2,\dots,x'_k) \mid x_ix'_i \in A(G_i), 1 \le i \le k\}$. Let $H^k = H \times H \times \dots H$, $k$ times.  

\begin{definition}[polymorphism, conservative polymorphism]
For a digraph $H$, a \emph{polymorphism} $\phi$ of arity $k$ on $H$ is a homomorphism from $H^k$ to $H$. 
For a polymorphism $\phi$,  $\phi(a_1,a_2,\dots,a_k)\phi(b_1,b_2,\dots,b_k)$ is an arc of $H$ whenever
$(a_1,a_2,\dots,a_k)(b_1,b_2,\dots,b_k)$ is an arc of $H^k$. We say $\phi$ is \emph{conservative} if $\phi(a_1,a_2,\dots,a_k) \in \{a_1,a_2,\dots,a_k\}$ for every $a_1,a_2,\dots,a_k \in V(H)$. 
\end{definition}



\begin{definition}[Semilattice, Majority, Maltsev polymorphisms]
A binary polymorphism $f$ on digraph $H$ is called \emph{semilattice} if $f(a,b)=f(b,a)$, and $f(a,a)=a$, $f(a,f(a,c))=f(f(a,b),c)$ for every $a,b,c \in V(H)$. A ternary polymorphism $g$ on $H$ is \emph{majority} if $g(a,a,b)=g(a,b,a)=g(b,a,a)=a$.  A polymorphism $h$ of arity $3$ is \emph{Maltsev} if for every $a,b \in V(H)$, 
 $h(a, b, b) = h(b,b, a) = a$. 
\end{definition}

\noindent The problem we consider is the generalization of the digraph list homomorphism problem.
\begin{definition}[list homomorphism]\label{list-definition}
  Let $G$ and $H$ be digraphs, and $L:G \rightarrow 2^H$ be a 
  set of lists.  The \emph{list homomorphism problem} for $H$, LHOM($H$), asks if there exists a   
  homomorphism $f$ from $G$ to $H$ such that (i) $\forall uv \in A(G), f(u) f(v) \in   
  A(H)$ (adjacency property) and (ii) $\forall u\in G, f(u) \in L(u)$ (list  property).
\end{definition}

It is known that if $H$ admits a Maltsev polymorphism then the core of $H$ is an induced directed path or is an induced directed cycle and hence the HOM($H$) is polynomial time solvable \cite{catarina}.

Our algorithm remains correct even if $H$ does not admit a Maltsev polymorphism but does admit
what we call a Maltsev \emph{list} polymorphism on $G \times H^3$ respecting $L$.


\begin{definition}[list polymorphism]
	Given digraphs $G, H$, and $L:G\rightarrow 2^H$,  $h$ is a \emph{list polymorphism} of arity $k$ on $G \times H^k$ 
    respecting the lists $L$ if 
    \begin{itemize}
        \item [(i)] $h: G \times H^k \rightarrow H$,  a homomorphism from $G \times H^k$ to $H$, i.e., $\forall x, y \in G$ and $a_1, ..., a_k\in L(x), b_1, ..., b_k\in L(y)$, if $xy\in G$, 
    $a_1~b_1\in A(H)$, .., $a_k~b_k\in A(H)$, then $h(x; a_1, ..., a_k)~h(y; b_1, ..., b_k)\in A(H)$.
    \item [(ii)] $\forall x\in G$ and $a_1, ..., a_k \in L(x)$, $h(x; a_1, ..., a_k)\in L(x)$.
    \end{itemize}
     In this situation, we say $L$ admits a polymorphism $h$ on $G \times H^k$, i.e. $h$ is a polymorphism on $G \times H^k$ respecting the lists $L$.  As the list is key, we denote the list homomorphism $h$ by $G \times_L H^k \rightarrow H$, and say $h$  is a list polymorphism on
     $G \times_L H^k$. 
    
    Polymorphism $h$ on $G \times_L H^3$ is called \emph{Maltsev} if for every $x\in G, a, b\in L(x)$, (i) $h(x; a,a,a) = a$, (ii)
    $h(x; a, b, b) = h(x; b, b, a) = a$. 
    \label{def:list-polymorphism}
\end{definition}
Note that if $H$ admits a Maltsev polymorphism $h$ of the normal kind, then for any $G$, $h$ yields a trivial list polymorphism $h'$ on $G \times H^3$ respecting lists $L$, for $L$ such that $L(x) = V(H)$ for all $x\in V(G)$
(by setting $h'(x; a_1, a_2, a_3) = h(a_1, a_2, a_3)$ for every $x\in G; a_1, a_2, a_3 \in L(x)$).

\begin{figure}[ht]
  \begin{center}
   \includegraphics[scale=0.55]{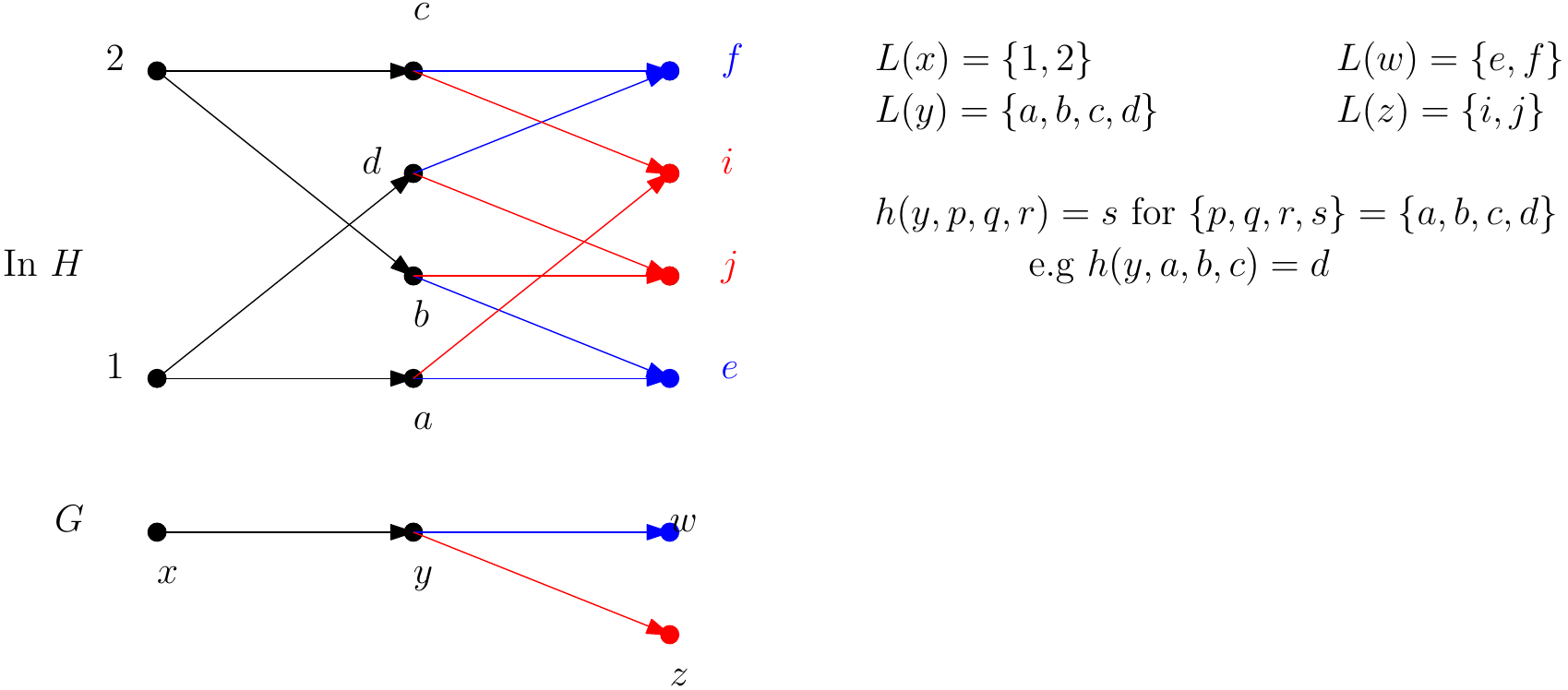}
  \end{center}
  \caption{ \small{Example of an instance that admits a Maltsev list polymorphism from $G$ to $H$, but where $H$ does
    not admit a majority polymorphism.  See Example \ref{ex:Maltsev-No-Majority}.}}
  \label{fig:Maltsev-No-Majority}
\end{figure}

However, the converse is not true -- that a Maltsev list polymorphism implies the existence of a Maltsev polymorphism of the usual kind. In the Figure \ref{fig:Maltsev-No-Majority}, there exists a Maltsev list polymorphism on $G \times H^3$, but $H$ itself does not admit a Maltsev (in the figure, for $H$ to have a Maltsev the arc $aj$ must be present).   

The polymorphism $g$ on $G \times_L H^3$, is called majority if for every $x\in G, a, b\in L(x)$, $g(x; a, b, b) = g(x; b, b, a) =g(x;b,a,b)= b$.  
It is known that in the usual setting, if digraph $H$ admits a Maltsev
polymorphism then it also admits a \emph{majority polymorphism}  \cite{kazda}. However, we note with the following
example that there exist $G, H, L$ such that (i) $G \times_L H^3$ admits a Maltsev  but (ii) $G \times_L H^3$ does not admit a majority list polymorphism  (and therefore also does not admit a majority polymorphism in the usual sense, or indeed a Maltsev polymorphism in the usual sense). 

\begin{example}
  Let $G$ and $H$ be as in Figure \ref{fig:Maltsev-No-Majority} so that $V(G)= \{x,y,z,w\}$, $V(H) = \{1,2,a,b,c,d,e,f,i,j\}$, and the edge sets and lists of $G$ and $H$ are as in the figure.  
  \label{ex:Maltsev-No-Majority}
\end{example}

\noindent $h$ as indicated in the figure is a Maltsev  polymorphism on $G  \times_L H^3$.  Suppose there exists a majority list polymorphism $g$ for $H$. Now $g(x;1,2,2)g(y;a,b,c) \in A(H)$, and hence, $g(y;a,b,c) \in \{b,c\}$. Moreover, 
$g(y;a,b,c)g(w;e,e,f) \in A(H)$ and hence $g(y;a,b,c) \in \{a,b\}$. These would imply that $g(y;a,b,c)=b$. On the other hand,  
$g(y;a,b,c)g(z,i,j,i) \in A(H)$, and hence, $g(y;a,b,c) \in \{a,c\}$, a contradiction. We conclude that $G \times_L H^3$ does not admit a majority list polymorphism.

\subsection{From CSP to Hypergraph List Homomorphism to Graph List Homomorphism }\label{CSP-to-Graphs}
Let $A$ be a finite set. By a $k$-ary relation $R$ on set $A$ we mean a subset of the $k$-th cartesian power $A^k$; $k$ is said to be the \emph{arity}
of the relation. A constraint language (relational structure) $\Gamma$ over $A$ is a set of relations over $A$. A constraint language is finite if it contains finitely many relations. Here we consider the finite languages.  

A hypergraph $\mathcal{G}$ on set $X$, consists of a set of hyperedges where each hyperedge $e$ is an {\em ordered} tuple $(x_1,x_2,\dots,x_k)$ , $x_1,x_2,\dots,x_k \in X$. Here $k$ is called the size of the hyperedge $e$. Notice that different hyperedges could have different sizes. 
A hypergraph is called uniform if all its hyperedges have the same size.  We denote the vertices of the hypergraph $\mathcal{G}$ by $V(\mathcal{G})$. 

For two hypergraphs $\mathcal{G},\mathcal{H}$, a homomorphism $f : \mathcal{G} \rightarrow \mathcal{H}$, is a mapping from $V(\mathcal{G})$ to $V(\mathcal{H})$ such that for every hyperedge $(x_1,x_2,\dots,x_k) \in \mathcal{G}$, $(f(x_1),f(x_2),\dots,f(x_k))$ is a hyperedge in  $\mathcal{H}$.

An instance of the constraint satisfaction problem (CSP) can be viewed as an instance of the hypergraph list homomorphism problem. We are given two hypergraphs $\mathcal{G},\mathcal{H}$ together with lists $\mathcal{L}$ where each hyperedge $\alpha \in \mathcal{G}$ has a list of possible hyperedges (all with the same size as $\alpha$) in $\mathcal{H}$, denoted by $\mathcal{L}(\alpha)$. 
The goal is to find a homomorphism  $f : \mathcal{G} \rightarrow \mathcal{H}$ such that for every hyperedge $\alpha=(x_1,x_2,\dots,x_k) \in \mathcal{G}$, $(f(x_1),f(x_2),\dots,f(x_k)) \in L(\alpha)$. In other words, if we look at the vertices of $\mathcal{G}$ as variables, the vertices of $\mathcal{H}$ as values, and hyperedges $\alpha's \in  \mathcal{G}$ as constraints then the existence of homomorphism $f$ illuminates a way of giving each variable a value, so that all constraints are satisfied simultaneously. A constraint is of form $(\alpha,L(\alpha))$, and a constraint is satisfied if tuple $\alpha$ is mapped by $f$ into one of the tuples in its list.

\begin{definition}[Signature]
    For every two hyperedges $\alpha_1,\alpha_2$ from $\mathcal{G}$ ( or $\mathcal{H}$) we associate a signature 
    $S_{\alpha_1,\alpha_2}=\{(i,j) | \text{ $\alpha_1[i]=\alpha_2[j]$ } \}$ ( $\alpha_1[i]$ is the element in coordinate $i$-th of $\alpha_1$).  
  
\end{definition} 

Let $\mathcal{H}$ be a hypergraph on set $A$. Let $\mathcal{H}_1,\mathcal{H}_2,\dots,\mathcal{H}_t$ be a partition of $\mathcal{H}$ into $t$ uniform hypergraphs. 
A mapping $h : A^r \rightarrow A$ is a polymorphism of arity $r$ on $\mathcal{H}$ if $h$ is closed under each $\mathcal{H}_i$, $1 \le i \le t$. In other words, for every $r$ hyperedges $\tau_1,\tau_2,\dots,\tau_r \in \mathcal{H}_i$, $h(\tau_1,\tau_2,\dots,\tau_r) \in \mathcal{H}_i$. Notice that here $h$ is applied coordinate wise (e.g. if $(a_1,a_2,a_3),(b_1,b_2,b_3),(c_1,c_2,c_3)$ are hyperedges in $H_i$ then $(h(a_1,b_1,c_1),h(a_2,b_2,c_2),h(a_3,b_3,c_3))$ is a hyperedge in $H_i$). \\

The following theorem translates the hypergraph list homomorphism problem (CSP) to an instance of the graph list homomorphism problem. 
\begin{theorem} \label{CSP-to-ListHOM}
Let $\mathcal{G},\mathcal{H},\mathcal{L}$ be an instance of hypergraph list homomrphism problem (CSP). Then there exist graphs $G,H$ with lists $L$ such that :
\begin{enumerate}

\item There is an $\mathcal{L}$-homomorphism from $\mathcal{G}$ to $\mathcal{H}$ if and only if there is an $L$-homomorphism from $G$ to $H$. 

\item If there exists a polymorphism $h$ of arity $r$ on $\mathcal{H}$ consistent with the lists $\mathcal{L}$ ( $h$ is closed in each $\mathcal{L}(\alpha)$, $\alpha \in \mathcal{G}$ ) then there exists a polymorphism $\phi$ on $G \times_L H^r$.  

\item If there exists a Maltsev polymorphism $h$ on $\mathcal{H}$ consistent with the lists $\mathcal{L}$ ( $h$ is closed in each $\mathcal{L}(\alpha)$, $\alpha \in \mathcal{G}$ ) then there exists a Maltsev polymorphism $\phi$ on $G \times_L H^3$.
\end{enumerate}
\end{theorem}
\pf We construct a list homomorphism instance $G,H,L$ as follows. \\
\begin{enumerate}
    \item For each hyperedge $\alpha \in \mathcal{G}$  consider a new vertex $\overline{\alpha}$ in graph $G$.
    \item For each hyperedge $\tau$ in $\mathcal{L}(\alpha)$ we consider a new vertex $\overline{\tau}$ in graph $H$.
    \item The list of $\overline{\alpha} \in G$ is  
$L(\overline{\alpha})=\{\overline{\tau} | \text{ $\tau \in \mathcal{L}(\alpha)$  } \}$.

\item There is an edge in $G$ between $\overline{\alpha}$ and $\overline{\beta}$ if $S_{\alpha,\beta} \ne\emptyset$.

\item There is an edge from vertex $\overline{\tau} \in L(\overline{\alpha})$ to $\overline{\omega} \in L(\overline{\beta})$ if $\overline{\alpha}\overline{\beta}$ is an edge of $G$, 
and $S_{\alpha,\beta} \subseteq S_{\tau,\omega}$. 

\end{enumerate}

The signature of edge $\overline{\alpha}\overline{\beta}$ in $G$ is defined to be  $S_{\alpha,\beta}$, and the signature of edge $\overline{\tau}\overline{\omega}$ in $H$ is defined to be $S_{\tau,\omega}$. 


\begin{claim}
There is an $\mathcal{L}$-homomorphism from $\mathcal{G}$ to $\mathcal{H}$ if and only if there is an $L$-homomorphism from $G$ to $H$. 
\end{claim}
Suppose there exists an $\mathcal{L}$-homomorphism $f : \mathcal{G} \rightarrow \mathcal{H}$ that maps each hyperedge $\alpha \in \mathcal{G}$ to  $\mathcal{L}(\alpha)$. Now we define a mapping $g : G \rightarrow H$ as follows: set $g(\overline{(x_1,x_2,\dots,x_k)})=\overline{(f(x_1),f(x_2),\dots,f(x_k))}$. 
We show that $g$ is a homomorphism. Consider two hyperedges $\alpha=(x_1,x_2,\dots,x_k)$ and $\beta=(y_1,y_2,\dots,y_{\ell})$ in $\mathcal{G}$, where $\overline{\alpha} \overline{\beta}$ is an edge of $G$. Thus, $S_{\alpha,\beta} \ne\emptyset$. 
Let $\tau=(f(x_1),f(x_2),\dots,f(x_k))$ and $\omega=(f(y_1),f(y_2),\dots,f(y_{\ell}))$. Since $f$ is an $\mathcal{L}$- homomorphism, we have $\tau \in \mathcal{L}(\alpha)$, and $\omega \in \mathcal{L}(\beta)$. This means that $\overline{\tau} \in L(\overline{\alpha})$ and $\overline{\omega} \in L(\overline{\beta})$. Suppose $\alpha[i]=\beta[j]$. Since $f$ is a homomorphism, $\tau[i]=\omega[j]$, and hence, $S_{\alpha,\beta} \subseteq S_{\tau,\omega}$.  Therefore, by the construction of $H$, $g(\overline{\alpha})g(\overline{\beta})$ is an edge in $H$, implying that $g$ is a list homomorphism from $G$ to $H$. 

Conversely, suppose $g : G \rightarrow H$ is an $L$-homomorphism. Define the mapping $f: V(\mathcal{G}) \rightarrow V(\mathcal{H})$ as follows. For vertex $\overline{\alpha} \in G$ with the corresponding hyperedge
$\alpha=(x_1,x_2,\dots,x_k)$, define $f(x_i)=b_i$, $1 \le i \le k$ where $g(\overline{(x_1,x_2,\dots,x_k)})=(b_1,b_2,\dots,b_k)$. 
We need to show that for every element $x \in V(\mathcal{G})$, $f(x)$ is uniquely defined. Suppose $x$ appears in two hyperedges, $\alpha=(x_1,x_2,\dots,x_i,x,x_{i+2},\dots,x_k)$ and $\beta=(y_1,y_2,\dots,y_j,x,y_{j+2},\dots,y_{\ell})$. Note that $\overline{\alpha}\overline{\beta}$ is an edge of $G$, and since $g$ is a list homomorphism, $\overline{\tau}=g(\overline{\alpha}) \in L(\overline{\alpha})$, and $\overline{\omega}=g(\overline{\beta}) \in L(\overline{\beta})$ are adjacent in $H$. Now according to the construction of $G,H,L$, this means $S_{\alpha,\beta}$ is a subset of $S_{\tau,\omega}$.  
Therefore, $\tau[i]=\omega[j]$, and hence, $f(x)$ gets only one unique value. By definition, it follows that for every hyperedge  $\gamma=(x_1,x_2,\dots,x_k) \in \mathcal{G}$, $g(\overline{\gamma}) \in L(\overline{\gamma})$, and hence,  
$(f(x_1),f(x_2),\dots,f(x_k))$ is in $\mathcal{L}(\gamma)$. \qed \\

\noindent {\em Proof of 2.} For every $r$, hyperedges $\tau_1,\tau_2,\dots,\tau_r$ from the the same list $\mathcal{L}(\alpha)$ we have $\tau=h(\tau_1,\tau_2,\dots,\tau_r) \in \mathcal{L}(\alpha)$ (here $h$ is applied coordinate wise). Now by definition we have $\overline{\tau} \in L(\overline{\alpha})$. 

Now define the mapping $\phi : G \times_L H^r \rightarrow H$ as follows. 
For every $\overline{\alpha} \in G$ and every $r$ hyperedges $\tau_1,\tau_2,\dots,\tau_r \in \mathcal{L}({\alpha})$,  $\overline{\tau_1},\dots,\overline{\tau_r} \in L(\overline{\alpha})$, set $\phi(\overline{\alpha}; \overline{\tau_1},\dots,\overline{\tau_r})=\overline
{h(\tau_1,\dots,\tau_r)}$. We show that $\phi$ preserve adjacency. Consider edge $\overline{\alpha}\overline{\beta}$ of $G$, and suppose $\overline{\tau_1},\dots,\overline{\tau_r} \in L(\overline{\alpha})$ and $\overline{\omega_1},\dots,\overline{\omega_r} \in L(\overline{\beta})$, where $\overline{\tau}_l \overline{\omega}_l \in E(H)$, $1 \le l \le r$. Then $\phi(\overline{x};\overline{\tau}_1,\dots,\overline{\tau}_r) \phi(\overline{y};\overline{\omega}_1,\dots,\overline{\omega}_r)$ is an edge of $H$. 

To see that we need to observe that by the construction of $H$, $S_{\alpha,\beta} \subseteq S_{\tau_l,\omega_l}$, $1 \le l \le r$. Therefore, if $a=\alpha[i]=\beta[j]$ for some $i,j$, 
we have $\tau_1[i]=\tau_2[i]=\dots=\tau_r[i]=a$ and $\omega_1[j]=\omega_2[j]=\dots =\omega_r[j]=a$. Let $\tau=h(\tau_1,\dots,\tau_r)$ and $\omega=h(\omega_1,\omega_2,\dots,\omega_r)$. Thus, $\tau[i]=a$ and $\omega[j]=a$, and hence, $(i,j) \in S_{\tau,\omega}$, and consequently $\overline{\tau}\overline{\omega} \in E(H)$.  
This would mean $\phi$ preserve adjacency and consequently is a polymorphism consistent with the lists $L$.  \\

\noindent {\em Proof of 3}. According to (2) for $r=3$, $\phi$ defined in (2) is a polymorphism on $G \times_L H^3$. Now it easy to see that 
$\phi(\overline{x};\overline{\alpha},\overline{\alpha},\overline{\alpha})= \overline{h(\alpha,\alpha,\alpha)}=\overline{\alpha}$, and $\phi(\overline{x};\overline{\alpha},\overline{\alpha},\overline{\beta})=
\overline{h(\alpha,\alpha,\beta)}=\overline{\beta}$,  
and $\phi(\overline{x};\overline{\beta},\overline{\alpha},\overline{\alpha})=
\overline{h(\beta,\alpha,\alpha)}=\overline{\beta}$. 

\qed

To prove Theorem \ref{maltsev-regontition}, it is enough to show that deciding whether a given hypergraph $\mathcal{H}$ admits a Maltsev polymorphism is polynomial time solvable. 

\begin{theorem}\label{Maltsev-hypergrap} 
Let $\mathcal{H}$ be a hypergraph. Then the problem of deciding whether $\mathcal{H}$ admits a Maltsev polymorphism is polynomial time solvable. 
\end{theorem}
\pf Let $\mathcal{H}_1,\dots,\mathcal{H}_k$ be the partitioned of $\mathcal{H}$ into uniform hypergraphs. 

We construct graph $G,H$ and lists $L$. The vertices of $G$ are triples $\overline{x}=(\alpha,\beta,\gamma)$ where $\alpha,\beta,\gamma \in \mathcal{H}_l$, $1 \le l \le k$. 
The vertices of $H$ are $\overline{\tau}$ where $\tau$ is a hyperedge of $\mathcal{H}$.  

For $\overline{x}=(\alpha,\beta,\gamma)$, $L(\overline{x})$, consists of all $\overline{\tau}$, $\tau \in \mathcal{H}_l$ satisfying the following conditions. 

\begin{itemize} 
\item If $\alpha[i]=\beta[i]=\gamma[i]=a$ then $\tau[i]$ is $a$.
\item  If $\alpha[i]=\beta[i]$ then $\gamma[i]=\tau[i]$.

\item If $\beta[i]=\gamma[i]$ then $\alpha[i]=\tau[i]$.
\end{itemize}

Two vertices $\overline{x}=(\alpha,\beta,\gamma)$, and $\overline{y}=(\alpha',\beta',\gamma')$ from $G$ with  $\alpha,\beta,\gamma \in \mathcal{H}_l$, $\alpha',\beta',\gamma' \in \mathcal{H}_t $  are adjacent if $S_{\alpha,\alpha'} \cap S_{\beta,\beta'} \cap S_{\gamma,\gamma'} \ne\emptyset$. 

Two vertices $\overline{\tau} \in L(\overline{x})$ and $\overline{\omega} \in L(\overline{y})$ in $H$ are adjacent if $S_{\alpha,\alpha'} \cap S_{\beta,\beta'} \cap S_{\gamma,\gamma'} \subseteq S_{\tau,\omega}$. Here $\overline{x}=(\alpha,\beta,\gamma)$, and $\overline{y}=(\alpha',\beta',\gamma')$.

\begin{claim}
$\mathcal{H}$ has a Maltsev polymorphism if and only if there is an $L$-homomorphism from $G$ to $H$.
\end{claim}
\pf Suppose $\mathcal{H}$ admits a Maltsev polymorphism. 
For every vertex $\overline{x}=(\alpha,\beta,\gamma) \in G$ where $\alpha,\beta,\gamma \in \mathcal{H}_l$, define mapping $g : G \rightarrow H$ with $g(\overline{x})=\overline{h(\alpha,\beta,\gamma)}$ , where $h$ is applied coordinate wise. Let $\overline{y}=(\alpha',\beta',\gamma')$ and suppose $\overline{x}\overline{y}$ is an edge of $G$. 
By definition, $\overline{\tau} \in L(\overline{x})$ where $\tau=h(\alpha,\beta,\gamma)$ and 
$\overline{\omega} \in L(\overline{y})$ where 
$\omega =h(\alpha',\beta',\gamma')$. Moreover, if  $(i,j) \in S_{\alpha,\alpha'},S_{\beta,\beta'},S_{\gamma,\gamma'}$ then the value of $i$-th coordinate of $\tau$ is $h(a_1,a_2,a_3)$ ($a_1,a_2,a_3$ are the $i$-th coordinate of $\alpha,\beta,\gamma$ respectively) and the value of the $j$-th coordinate of $\omega$ is $h(a_1,a_2,a_3)$ ($a_1,a_2,a_3$ are the $j$-th coordinate of $\alpha',\beta',\gamma'$ respectively). Therefore, $(i,j) \in S_{\tau,\omega}$, and hence, there is an edge from $\overline{\tau}$ to $\overline{\omega}$ in $H$. Therefore, $g$ is a homomorphism from $G$ to $H$. \\

Conversely, suppose $g$ is an $L$-homomorphism from $G$ to $H$. Suppose $\overline{\tau}=g(\overline{x})$ for $x=(\alpha,\beta,\gamma)$. Then, 
for every $a_1,a_2,a_3$ that are the $i$-th coordinate of $\alpha,\beta,\gamma$, respectively, set $h(a_1,a_2,a_3)=a_4$ where $a_4$ is the $i$-th coordinate of $\tau$ (recall $\tau$ is a ordered hyperedge  corresponding to $\overline{\tau}=g(\overline{x})$). 

Notice that if $a_1=a_2$ then because of the way we construct the lists, $h(a_1,a_2,a_3)=a_3$, and similarly when $a_2=a_3$, $h(a_1,a_2,a_3)=a_1$. Consider vertex $\overline{y} \in G$ with $\overline{y}=(\alpha',\beta',\gamma')$. Suppose the $j$-coordinate of $\alpha',\beta',\gamma'$ are $a_1,a_2,a_3$ respectively. Let $\overline{\omega}=g(\overline{y})$. By definition, $h(a_1,a_2,a_3)$ is $a'_4$ where $a'_4$ is the $j$-th coordinate of $\omega$. We show that $a_4=a'_4$. Observe that $(i,j) \in S_{\alpha,\alpha'} \cap S_{\beta,\beta'} \cap S_{\gamma,\gamma'}$ where $a_1 $ appears in $i$-th coordinate of $\alpha$ and in the $j$-th coordinate of $\alpha'$;  $a_2$ is an element  appearing in $i$-th coordinate of $\beta$ and in the $j$-th coordinate of $\beta'$; and finally  $a_3$ appears in the $i$-th coordinate of $\gamma$ and in the $j$-th coordinate of $\gamma'$.  
Therefore, $\overline{x},\overline{y}$ are adjacent in $G$, and since $g$ is a homomorphism, $\overline{\tau}$ and $\overline{\omega}$ must be adjacent in $H$. By the construction of the lists, the $i$-th coordinate of $\tau$ is the same as the $j$-th coordinate of $\omega$, i.e. $a_4=a'_4$. Notice that since $g$ is a list homomorphism, 
$\overline{\tau} \in L(\overline{x})$ where $\tau= h(\alpha,\beta,\gamma)$, and hence, $\tau$ belongs to $\mathcal{H}$. \qed \\

\begin{claim}
If $\mathcal{H}$ admits a Maltsev polymorphism then $G \times_L H^3$ admits a Maltsev polymorphism. 
\end{claim}
\pf 
For each vertex $\overline{x}=(\alpha,\beta,\gamma)$, let $\mathcal{L}(\alpha,\beta,\gamma)=\{ \tau \in \mathcal{H} \mid \overline{\tau} \in L(\overline{x}) \}$.
Now it is easy to observe that each list $\mathcal{L}(\alpha,\beta,\gamma)$ is closed under Maltsev polymorphims $h$. Indeed, if no pair of $\alpha,\beta,\gamma$ shares an element, then $\mathcal{L}(\alpha,\beta,\gamma)=\mathcal{H}_l$ and by definition $\mathcal{H}_l$ is closed under $h$. If $\alpha[i]=\beta[i]$ for some $i$ then all hyperedges in $\mathcal{L}(\alpha,\beta,\gamma)$ have $\gamma[i]$ in their $i$-th coordinate, and since the projection of $\mathcal{H}_l$ on the $i$-th coordinate still has Maltsev polymorphism $h$, $\mathcal{L}(\alpha,\beta,\gamma)$ is closed under $h$. Similarly when $\beta,\gamma$ or $\alpha,\gamma$ share elements, we conclude that $\mathcal{L}(\alpha,\beta,\gamma)$ is closed under $h$. 

By observation above for lists $\mathcal{L}$ and similar argument from the proof of  \ref{CSP-to-ListHOM} (3), it is easy to see that the existence of polymorphism $h$ on $\mathcal{H}$ implies the existence of a Maltsev polymorphism $f$ on $G \times_L H^3$. Notice that $f$ is closed under the list because $\mathcal{L} (\alpha,\beta,\gamma)$ is closed under Maltsev polymorphims $h$. \qed \\

By the above claim, $G \times_L H^3$ admits a Maltsev polymorphism, and hence, according to Theorem \ref{Maltsev-poly} finding a homomorphism from $G$ to $H$ with respect to lists $L$ is polynomial time solvable. Therefore, we can obtain a Maltsev polymorphism for $\mathcal{H}$ if one exists. \qed \\

The following theorem shows that finding a Maltsev polymorphism for a given relational structure $R$ is polynomial time solvable. 

\begin{theorem}\label{Maltsev-relational} 
Let $\mathcal{R}$ be a relational structure on set $A$. Then the problem of deciding whether $\mathcal{R}$ admits a Maltsev polymorphism is polynomial time solvable. 
\end{theorem}
\pf It is easy to see that $\mathcal{R}$ is just a hypergraph, and hence, by Theorem \ref{Maltsev-hypergrap} finding a Maltsev polymorphism for $\mathcal{R}$ (if one exists) is a polynomial time task. \qed

\section{Algorithm}

\paragraph{Preprocessing}\label{2-3-consistency}
One of the common ingredients in the $CSP$ algorithms is the use of consistency checks to reduce the set of possible values for each variable (see, for example the algorithm
 outlined in \cite{pavol-book}).
Our algorithm includes such a consistency check as a first step.  We keep \emph{pair lists}, $L(x, y)$ for each pair of vertices $x, y\in G$ where $(a, b) \in L(x, y)$ indicates that $a,b\in H$, $a\in L(x)$, $b\in L(y)$, and the algorithm considers a simultaneous assignment of $x\rightarrow a$ and $y\rightarrow b$ as possible (at the moment).  We ensure the pair lists are 
(2,3)-consistent \cite{FV93} to enforce the following.  
(i) \emph{Arc consistency} -- if $xy\in A(G)$ and $(a,b) \in L(x,y)$ then $ab\in A(H)$. (ii) \emph{Pair consistency} -- 
if $(a,b)\in L(x,y)$ and $z\in V(G)$ then there must exist $c\in L(z)$ with $(a,c)\in L(x,z)$ and $(c,b)\in L(z,y)$.  We call the procedure to enforce these conditions, \emph{Preprocessing}.


\begin{definition}[connected component ]
By \emph{connected component} of $G \times_L H$ we mean a \emph{weakly connected component} $C$ of digraph $G \times_L H$(i.e. a connected component of $G \times_L H$ when we ignore the direction of the arcs) which is closed under $(2,3)$-consistency. That means, for every $(x,a),(y,b) \in C$, and every $z \in G$ there is some $c \in L(z)$ such that $(a,c) \in L(x,z)$, $(b,c) \in L(y,z)$.  
\end{definition}


\paragraph{Algorithm main loop}
The algorithm takes $G, H, L$ as an input. It starts by Preprocessing the lists and if it counters an empty list then there is no homomorphism from $G$ to $H$. We assume the instance is connected; otherwise, the algorithm considers each connected component of $G \times_L H$ separately.  
Next at each step in the main loop, the algorithm considers a 
vertex $x \in G$ and two vertices $a,b \in L(x)$ with the goal of eliminating either $a$ or $b$ 
from $L(x)$. 

To decide whether to remove $a$ or $b$ it constructs a smaller instance of the 
problem with respect to $a$, say $(G',H,L')$, and solves this instance recursively (performing a smaller test). Here $G'$ is an induced sub-digraph of $G$ , and for every $y \in G'$, $L'(y)$ is the set of all the elements $e \in L(y)$ so that $(a,e) \in L(x,y)$. Notice that by definition, $L'(x)=\{a\}$. Now it is easy to see that we can remove $a$ form $L(x)$ if this instance does not have a solution because there is no homomorphism from $G$ to $H$ that maps $x$ to $a$. 

After all the smaller tests, if both $a,b$ remained in the list of $x$ then we remove $b$ from $L(x)$. To prove correctness we must also show that if the smaller instance have a solution then we can remove $b$ from $L(x)$ and maintain the presence of a homomorphism. At the end  we are left with singleton lists and check whether the singleton lists yeild a homomorphism from $G$ to $H$. 

An {\em oriented walk (path)} is obtained from a walk (path) by orienting each of its edges. 
An {\em oriented cycle} is obtained from a cycle by orienting each of its edges. 
Let $X = x_1,x_2,\dots,x_n$ be a walk in digraph $H$. When $x_ix_{i+1}$, $1 \le i \le n-1$, is an arc of $H$ then  $x_ix_{i+1}$ is called a forward arc, otherwise, (when $x_{i+1}x_i \in A(H)$), $x_ix_{i+1}$ is called a backward arc. We say oriented walk $Y=y_1,y_2,\dots,y_n$ is congruent to $X$ if they follow the same patterns of forward and backward arcs. In other words, for every $1 \le i \le n-1$, $x_ix_{i+1},y_iy_{i+1}$ are both forward or both are backward arcs. 

\begin{figure}
  \begin{center}
   \includegraphics[scale=0.5]{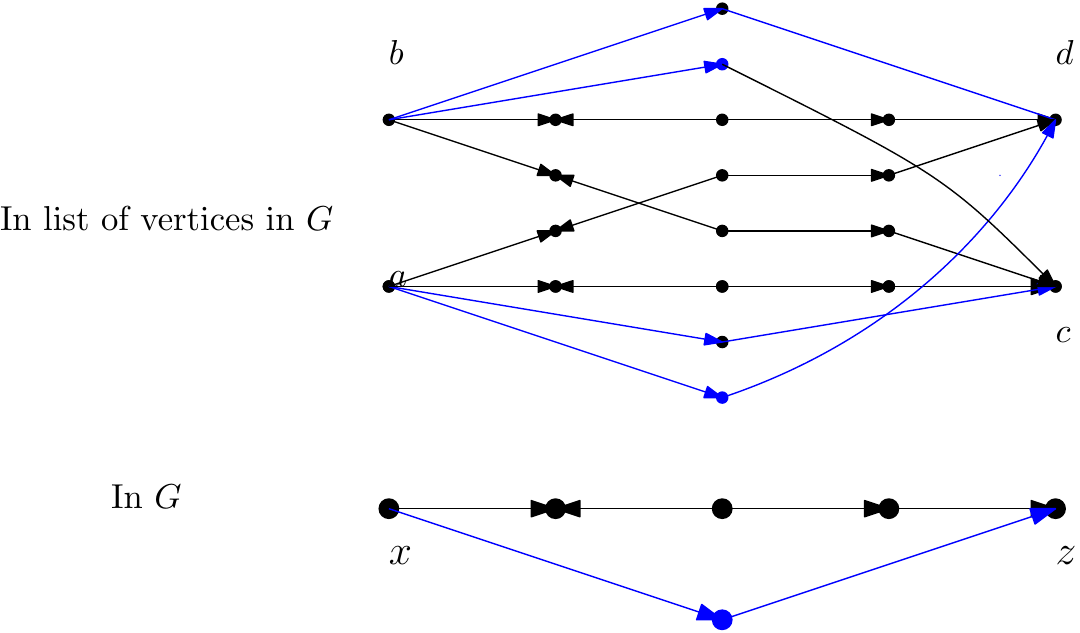}
  \end{center}
  \caption{Example $G$, $H$, and lists with rectangle property.  $(a,b)$ in $L(x)$ lies on a rectangle with $(c,d)$ in $L(z)$.  See Definition \ref{def:rectangle}.  }
\label{fig:rectangle-property}
 \end{figure}

\paragraph{Rectangle Property}
Algorithms for graphs that admit Maltsev polymorphisms often rely on the so-called rectangle property of these instances.  Adapted to the list homomorphism setting, we have Definition \ref{def:rectangle} and Lemma \ref{lm:rectangle}.   
\begin{definition}[rectangle property]
Let $G, H, L$ be an instance of the list homomorphism problem.  Two vertices $a,b \in L(x)$ lie on a {\em rectangle} if there exists $y \in V(G)$, and two distinct elements $c,d \in L(y)$ such that for any oriented path $Y$ from $x$ to $y$ in $G$, there exist congruent walks $A_1$ (from $a$ to $c$), $A_2$ (from $b$ to $d$), $B_1$ (from $a$ to $d$) and $B_2$ (from $b$ to $c$) all in $L(Y)$, i.e.  $(a,c),(a,d),(b,c),(b,d) \in L(x,y)$. In Figure \ref{fig:rectangle-property}, $a, b$ from $L(x)$ lies on a rectangle with $c,d$ in $L(z)$.
\label{def:rectangle}
\end{definition}

For an oriented path $X$ in $G$, let $L(X)$ denote the list of the vertices in $X$. 

\begin{lemma}[rectangle property]\label{Maltsev}
Let $G, H, L$ be an instance of the list homomorphism problem such that $G \times_L H^3$ admits a Maltsev  polymorphism $h$. 

Let $X$ be an oriented path in $G$ and let $B,C,D$ be three walks in $L(X)$ all congruent to $X$ where $B$ is from $a$ to $c$, 
 $C$ is from $b$ to $c$, and $D$ is from $b$ to $d$. Then there exists a walk $E$ from $a$ to $d$ in $L(X)$ which  is congruent
with $X$.
\label{lm:rectangle}
\end{lemma}
\pf Let $b_1, b_2, ..., b_\ell$ be the vertices in $B$ with $b_1=a$ and $b_\ell=c$ and similarly let $C: c_1,c_2,\dots,c_{\ell}$ (with $c_1=b$, $c_{\ell}=c$), $D: d_1,d_2,\dots,d_{\ell}$ (with $d_1=b$, $d_{\ell}=d$), and $X : x_1,x_2,\dots,x_{\ell}$. Because $h$ is a polymorphism we know that $E: h(x_1; b_1, b_1,d_1)$, ..., $h(x_\ell; b_\ell, b_\ell, b_\ell)$ is a walk in $L(X)$ that is congruent to $X$. By the definition of the walks and the Maltsev property of $h$, $h(x_1; b_1, c_1, d_1) = h(x_1; a,b,b) = b = c_1$, and $h(x_\ell; b_\ell, b_\ell, d_\ell)=h(x_\ell; c,c,d)=d=d_\ell$.  $E$ is the walk claimed in the lemma.
\qed

\begin{observation}\label{obs1}
Let $\alpha,\beta,\gamma \in L(x)$ before Preprocessing. Then $h(x;\alpha,\beta,\gamma)$ stays in $L(x)$ after Preprocessing. 
\end{observation}
\pf
Since $a,b,c \in L(x)$ after Preprocessing, for every out-neighbor (in-neighbor) $y$ of $x$, there exist $a',b',c' \in L(y)$ (after Preprocesssing) such that $aa' , bb',cc' \in A(H)$.  Therefore, by definition \\ $h(x;a,b,c) h(y;a',b',c') \in A(H)$, and hence, we conclude $h(x;a,b,c) \in L(x)$.
\qed

Now Lemma \ref{Maltsev} implies the following corollary. 
\begin{corollary} \label{all-in}
Let $G, H, L$ be an instance of the list homomorphism problem such that $G \times_L H^3$ admits a Maltsev polymorphism $h$.  
Suppose after Preprocessing $(a,c),(b,c),(b,d) \in L(x,y)$. Then $(a,d) \in L(x,y)$.
\end{corollary}
\pf If $x$ and $y$ are disconnected in $G$ then the statement is trivially true.  Otherwise, let $W$ be a walk from $x$ to $y$ in $G$. $(a,c)\in L(x,y)$ after Preprocessing implies the existence of a walk $AC$ congruent to $W$ from $a$ to $c$ in $L(W)$.  Similarly, $(b,c),(b,d)\in L(x,y)$ imply the existence of a walk $BC$ (from $b$ to $c$), and a walk $BD$ (from $b$ to $d$) in $L(W)$ that are congruent to $W$.

By Observation \ref{obs1}, the existence of a Maltsev polymorphism with respect to the new lists, is preserved. Therefore, by applying Lemma \ref{Maltsev}, there exists a walk in $L(W)$ from $a$ to $d$ which is congruent to $W$. Therefore $(a,d) \in L(x,y)$. \qed

We note that in the usual setting the existence of a Maltsev polymorphism for $H$ is equivalent to the rectangle property holding for $H$.  In the list setting this is not true in general, with the next example giving a counter example.

\begin{figure}
  \begin{center}
   \includegraphics[scale=0.5]{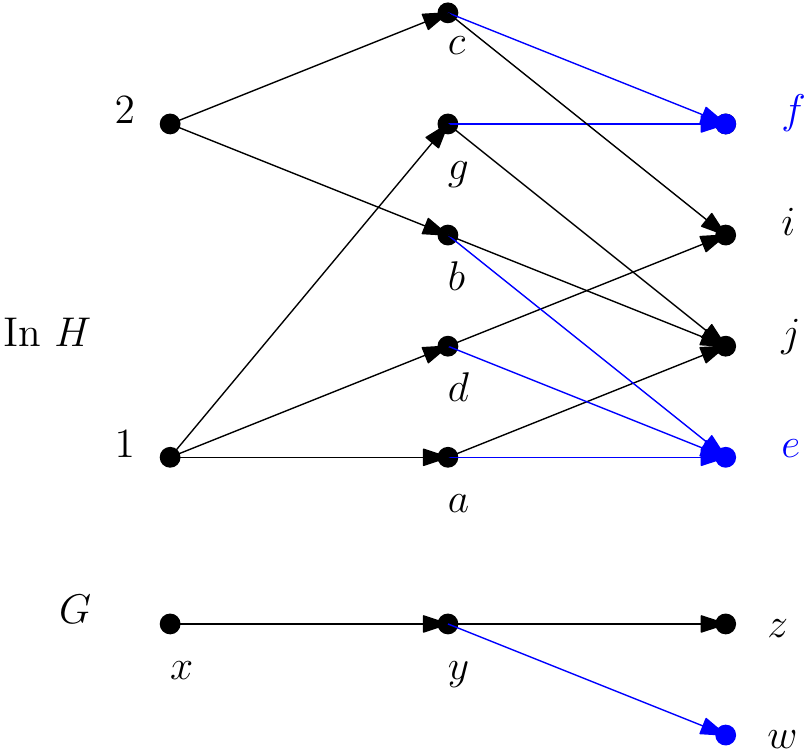}
  \end{center}
  \caption{Example $G$ and $H$ such that the rectangle property of Definition \ref{def:rectangle} holds, but there is no Maltsev list polymorphism from $G$ to $H$.  Here we have lists: $L(x) = \{1, 2\}, L(y) = \{a,b,c,d,g\}, L(z)=\{i,j\}, L(w)=\{e,f\}$.  See Example \ref{ex:Rectangle-No-Maltsev}. }
\label{fig:Rectangle-No-Maltsev}
 \end{figure}

\begin{example}
  Consider $G, H, L$ from Figure \ref{fig:Rectangle-No-Maltsev}.  These satisfy the rectangle property (Definition \ref{def:rectangle}) but do not admit a Maltsev list polymorphism.
  \label{ex:Rectangle-No-Maltsev}
\end{example}
\noindent It can be verified that the example satisfies the rectangle property (Definition
\ref{def:rectangle}) for all pairs of vertices. However, there cannot be a Maltsev list
polymorphism consistent with the lists given in the figure, as follows.  If $h$ were a Maltsev list polymorphism then because $yz \in A(G)$, $aj,bj,ci \in A(H)$, we must have : $h(y;a,b,c)h(z;j,j,i) \in A(H)$, and hence, $h(y;a,b,c)\in\{d,c\}$. Moreover, since $xy \in A(G)$, $1a,2b,2c \in A(H)$ we have $h(x;1,2,2)h(y;a,b,c) \in A(H)$, and hence, $h(y;a,b,c) \in \{a,d,e\}$. Therefore, $h(y;a,b,c)=d$. On the other hand, since $yw \in A(G)$, $ae,be,cf \in A(H)$, we have  $(y;a,b,c)(w;e,e,f)$ as an arc of $G \times_L H^3$, implying  $h(y;a,b,c)h(w;e,e,f)$ to be an arc of $H$, and consequently, $df \in A(H)$, a contradiction. 

After performing Preprocessing, the algorithm  separates  $G\times_{L} H$ into weakly connected components.  So in what follows we may assume that $G\times_L H$ is a single weakly connected component. 

\begin{definition}
For vertex $x \in V(G)$ and $a \in L(x)$, let $L_{x,a}$ be the restriction of $L$ when the list of $x$ has only element $a$. In other words, for every $y \in V(G)$, 
$L_{x,a}(y)= \{ \text {$d \ \ | (a,d) \in L(x,y)$  } \}$. 

The pair list of $L_{x,a}(y,z)$ for $y,z \in V(G)$ is the set of $(c,d)$ that are $(2,3)$- consistent within $L_{x,a}$. In other words, $(c,d) \in L_{x,a}(y,z)$ when  $c \in L_{x,a}(y), d \in L_{x,a}(z)$, 
and for every $w \in V(G)$ there exists some $e \in L_{x,a}(w)$ such that  $(e,c) \in L_{x,a}(w,y)$, and 
$(e,d) \in L_{x,a}(w,z)$. 
\end{definition}

We say two distinct vertices $a,b$ in $L(x)$ are {\em twins} if for every in-neighbor (out-neighbor) $y$ of $x$, the in-neighborhood (out-neighborhood) of $a,b$ in $L(y)$ are the same. It is clear in this case we can remove one of $a,b$ from $L(x)$. So we assume we get ride of twin vertices first because they are identical. We further extend this notion to a bigger sub-digraph. 

\begin{definition}[Identical Vertices]\label{identical}
We say two vertices $a,b \in L(x)$ are {\textbf {identical with respect to set $B \subseteq V(G)$}}  if 
\begin{itemize}
    \item [(1)] For every $y \in B$, $L_{x,a}(y)=L_{x,b}(y)$. 
    \item [(2)] For every $y,z \in B$, and every $c,d \in V(H)$,  
    $(c,d) \in L_{x,a}(y,z)$ if and only if $(c,d) \in L_{x,b}(y,z)$. 
\end{itemize}

\end{definition}

The construction of $G'$ for the smaller test starts by looking at sub-digraph $G^{L,x}_{a,b}$ which essentially is the union of all the minimal rectangles with one side being $x,a,b$. The following is the formal definition of $G^{L,x}_{a,b}$.  

\begin{definition}[$G^{L,x}_{a,b}$]\label{GXab}
For $a,b \in L(x)$, initially $G^{L,x}_{a,b}$ is the induced sub-digraph of $G$ with vertices $y$ such that $L_{x,a}(y) \setminus L_{x,b}(y) \ne \emptyset$. 
Let $B_1$ be the set of vertices of $G \setminus G^{L,x}_{a,b}$ with out-neighbor (in-neighbor) to a vertex in $G^{L,x}_{a,b}$. 
Add the vertices of $B_1$ into $G^{L,x}_{a,b}$ together with their connecting arcs. We call the set $B_1$, the 
{\textbf{boundary}} vertices in $G^{L,x}_{a,b}$ and denote it by $B(G^{L,x}_{a,b})$. Note that by rectangle property, 
$B(G^{L,x}_{a,b})=B(G^{L,x}_{b,a})$.

\end{definition}

\begin{figure}
  \begin{center}
   \includegraphics[scale=0.7]{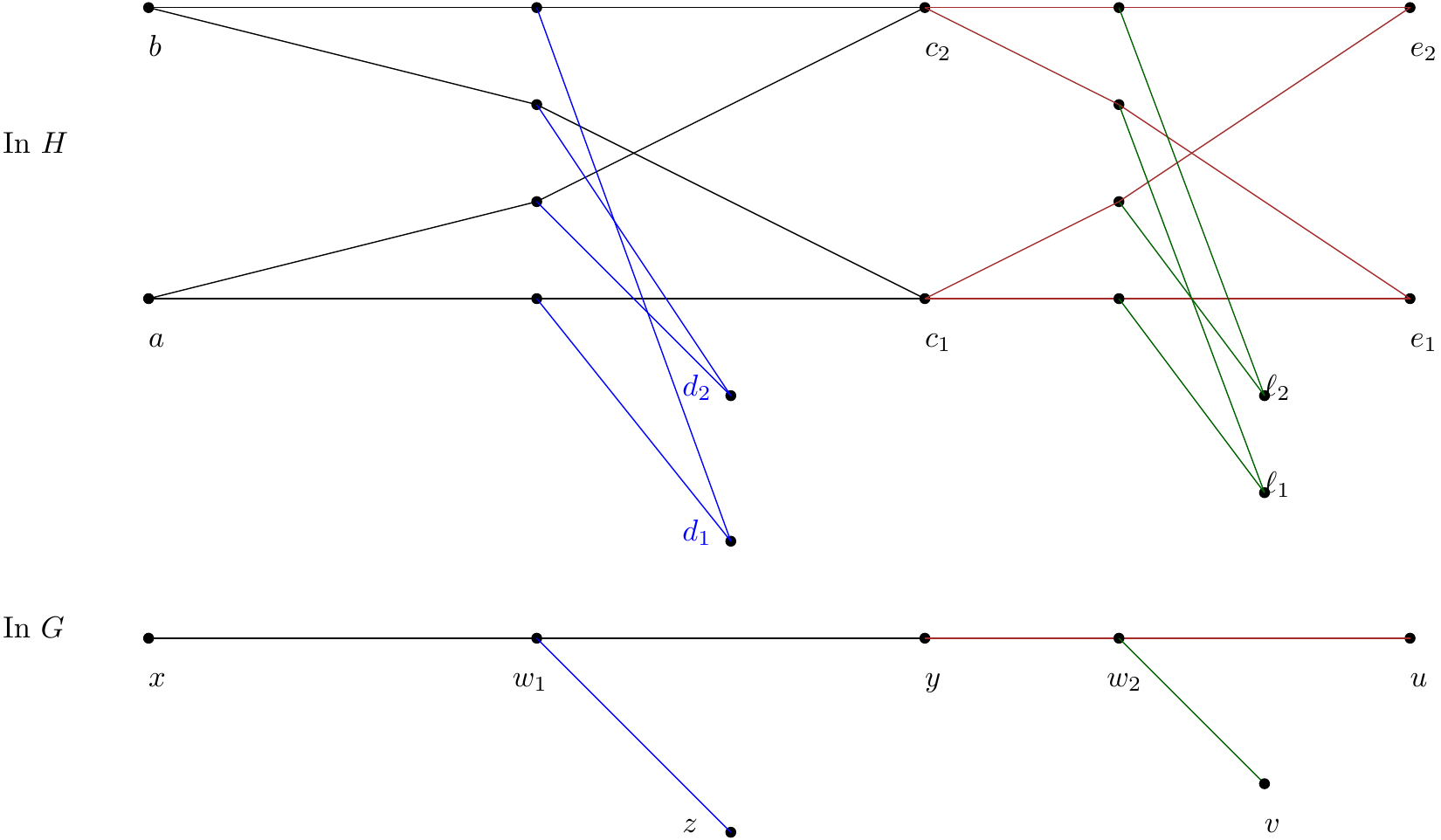}

  \end{center}
  \caption{ \small{Pair lists are different in $L_{x,a}(y,z)$, and $L_{x,b}(y,z)$ }}
\label{fig:not-identical}
 \end{figure}

If $a,b \in L(x)$ are not identical w.r.t. $B(G^{L,x}_{a,b})$, then there exist $y,z \in B(G^{L,x}_{a,b})$ and $c_1,c_2 \in L_{x,a}(y)$ and $d_1,d_2 \in L_{x,a}(z)$ such that $(c_1,d_1) ,(c_2,d_2)\in L_{x,a}(y,z)$ but $(c_2,d_1),(c_1,d_2) \not\in L_{x,a}(y,z)$, and $(c_2,d_1) ,(c_1,d_2)\in L_{x,b}(y,z)$, but $(c_1,d_1),(c_2,d_2) \not\in L_{x,b}(y,z)$.
We say $y$ and $c_1,c_2 \in L_{x,a}(y)$ witness $x,a,b$ 
 at $z,d_1,d_2$. Similarly we  say $z,d_1,d_2$ witness $x,a,b$ at $y,c_1,c_2$ (see Figure \ref{fig:not-identical}).

\paragraph{An overview of the main part  of the algorithm}

The core of the Algorithm is the construction of the smaller instance, i.e. the induced subgraph;  $\widehat{G^{L,x}_{a,b}}$ of $G$. This is done in function \Call{Sym-Dif}. We start with $G^{L,x}_{a,b}$ according to the definition \ref{GXab}. If $a,b \in L(x)$ are identical then 
$G'=\widehat{G^{L,x}_{a,b} }=G^{L,x}_{a,b} \setminus B(G^{L,x}_{a,b})$, and $L'=L_{x,a}$. 


If $a,b \in L(x)$ are not identical then the algorithm continues adding other vertices into $G'$. 
The reason is that, if the instance $G^{L,x}_{a,b},H,L'$ has a solution, then by removing $b$ from $L(x)$ we may loose the optimal solution, and this is because of the existence of a vertex $y$ in the boundary of $G^{L,x}_{a,b}$, say $B'$, and two vertices $c_1,c_2 \in L(y)$ that witness $x,a,b$. The \Call{Sym-Dif}{}, adds  the vertices of $G^{L,y}_{c_1,c_2} \setminus B(G^{L,y}_{c_1,c_2})$ into $G'$. The boundary vertices of $G^{L,y}_{c_1,c_2}$ are added into $B'$;  and we remove the vertices from $B'$ that are in $G^{L,y}_{c_1,c_2} \setminus B(G^{L,u}_{c,d})$. Now the recent frontier for identical measure are $y,c_1,c_2$ with respect to the new boundary. If there are no non-identical pairs left then we stop, otherwise, we continue finding new witnesses and grow $G'$ accordingly.  
In the Figure \ref{fig:not-identical} we proceed by adding $u,v$ as the boundary vertices of the $\widehat{G^{L,x}_{a,b}}$. $\widehat{G^{x,L}_{a,b}}$ has vertices $x,w_1,w_2,y$.

\begin{algorithm}
  \caption{RemoveMinority -- Using Maltsev Property }
 
  \label{alg-remove-minority-new}
  \begin{algorithmic}[1]
  \Function{RemoveMinority}{$G,H,L$}

  \State \Call {Preprocessing}{$G,H,L$} and if a list becomes empty then {\textbf {return}} $\emptyset$ 
  
  \State Consider each connected component of $G \times_L H$ separately
  
  \Comment {we assume $G \times_L H$ is connected} 
  
  \State $\forall \ \ x \in V(G)$ and $\forall a,b \in L(x)$, if $a,b$ are twins then remove $b$ from $L(x)$.

   
  
 \ForAll{ $x \in V(G), a,b \in L(x)$ with $a \ne b$ }
 
   \State $(\widehat{G^{L,x}_{a,b}},L')=$ \Call{Sym-Dif}{$G,H,L,x,a,b$} \label{line5}
   
   \State $g^x_{a,b}=$ \Call{RemoveMinority}{$\widehat{G^{L,x}_{a,b}},H,L'$} \label{line6}
    \If{  $g^x_{a,b}$ is empty } remove $a$ from $L(x)$ \label{line7}
    
    \Else {} remove $b$ from $L(x)$ \label{line8}

    \EndIf
    
    \State  Preprocessing ($G,H,L$)
 \EndFor 
  
 

\State Set $\psi$ to be an empty homomorphism.

 \If { $\exists x \in V(G)$ with $L(x)$ is empty }  return $\emptyset$. 
  \Else {} 
 \ForAll{ $x \in V(G)$ }  
 \State  $\psi(x)= L(x)$  \Comment{in this case the lists are singletons } 

 \EndFor

\EndIf

\State {\textbf {return} } $\psi$

\EndFunction

\Statex

\Function{Sym-Dif}{$G,H,L,x,a,b$}

\State Set $\widehat{G^{L,x}_{a,b}}=G^{L,x}_{a,b} \setminus B(G^{L,x}_{a,b})$, and $B'=B(G^{L,x}_{a,b})$ 

\State Set $S$ to be empty  \Comment{ $S$ is a stack}

\State Set $L_1= L_{x,a}$
\State push $x,a,b$ into $S$

\While{ $S$ is not empty }

\State pop $(x',a',b')$ from $S$


\ForAll{ $u \in B'$ and $c_1,c_2 \in L_1(u)$  s.t. $u,c_1,c_2$ witness $x',a',b'$ at $v,d_1,d_2$
}


\State Add new vertices from $G^{L_1,u}_{c_1,c_2} \setminus B(G^{L_1,u}_{c_1,c_2})$ into $\widehat{G^{L,x}_{a,b}} $

\State Update $B'$ by adding new boundary vertices from $B(G^{L_1,u}_{c_1,c_2})$ and removing the old 

\hspace{14mm} boundary 
vertices  from $B'$ that become internal vertices. 

\State push $(u,c_1,c_2)$ into $S$

\EndFor


\EndWhile

\State Initialize new lists $L'$

\State $\forall y \in \widehat{G^{L,x}_{a,b}}$, set $L'(y)=L_{x,a}(y)$ 
\Comment{in the next call we make sure the $L'$ lists are $(2,3)$-consistent}


\State Set $B(\widehat{G^{L,x}_{a,b}})=B'$ 
\Comment{this setting is for referring in the proof} \label{line31}

\State {\textbf{return}} $(\widehat{G^{L,x}_{a,b}},L')$ 

\EndFunction

\end{algorithmic}
\end{algorithm} 

\newpage
\paragraph{Example, System of linear equations:}
\begin{figure}
  \begin{center}
   \includegraphics[scale=0.7]{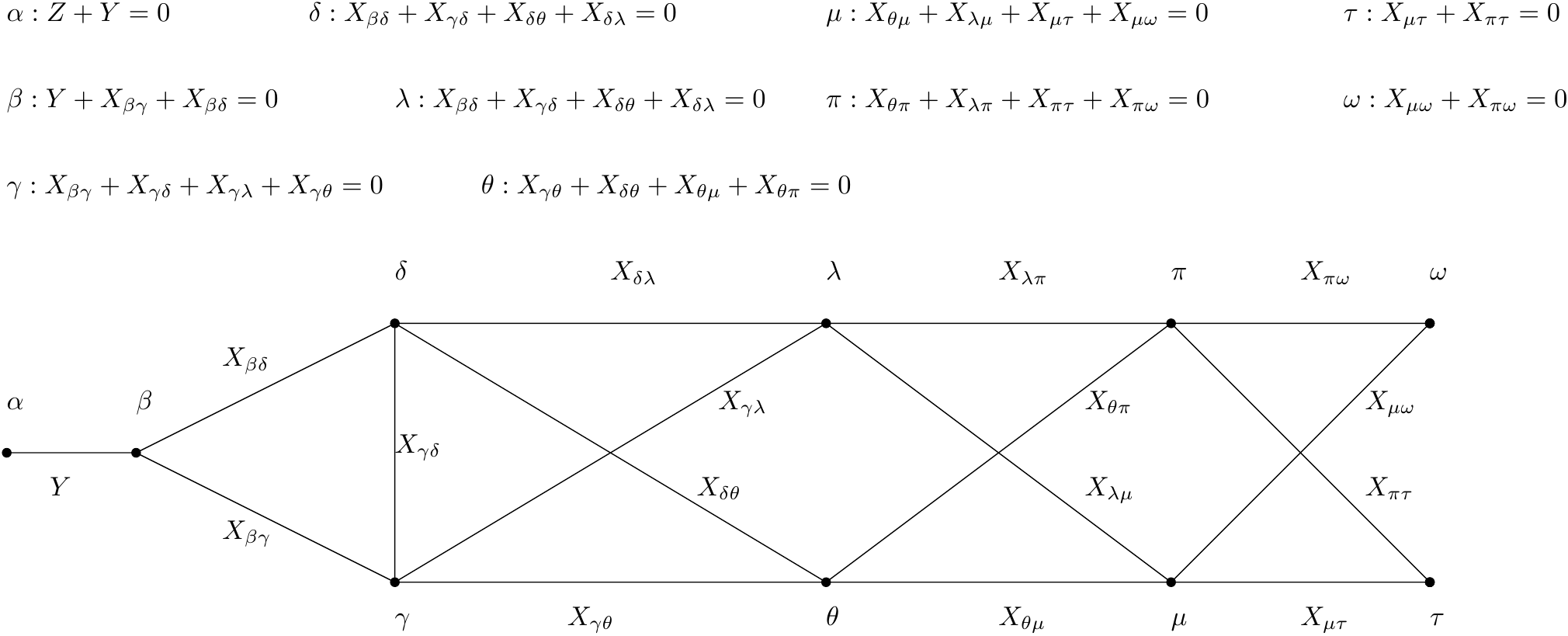}

  \end{center}
  \caption{\small{ The graph corresponding to the equation with unique  solution} }
\label{fig:equations}
 \end{figure}
Consider the following system of linear equations in $Z_2$.
This system of equations has a solution in $Z_2$. For $Y=Z=1$, the system does not have a solution because of the parity reason and when $Y=Z=0$ the system has several solutions. Each equation is considered as a vertex of the graph $G$ depicted in Figure \ref{fig:equations}. Two vertices are adjacent if they share a variable. The list of each vertex $v \in V(G)$; $L(v)$, consists of all the solutions for equation $v$. For example, $L(\alpha)=\{00,11\}=\{0,1\}$ where $0,0$ are the values that might be assigned to $Y,Z$ respectively. Here we index $00$ by $0$ and $11$ by $1$.   $L(\beta)=\{000,011,101,110\}=\{2,3,4,5\}$, $000 \in L(\beta)$ is given index $2$, $011$ index $3$ and so on.  $L(\gamma)=\{0011,0101,0110,1001,1010,1100,1111\}=\{6,7,8,9,10,11,12,13\}$, $L(\delta)=\{0011,0101,0110,1001,1010,1100,1111\}=\{14,15,16,17,18,19,20,21\}$,\\ $L(\theta)=\{0011,0101,0110,1001,1010,1100,1111\}=\{22,23,24,25,26,27,28,29\}$, \\
$L(\lambda)=\{0011,0101,0110,1001,1010,1100,  1111\}=\{30,31,32,33,34,35,36,37\}$, where for example, $1100$ represents a potential assignment for $X_{\gamma\lambda}=1, X_{\delta \lambda}=1,  X_{ \lambda \theta}=0, X_{\lambda \pi}=0$. 
$L(\mu)=\{0011,0101,0110,1001, 1010,1100,1111\}=\{38,39,40,41,42,43,44,45\}$,  $L(\pi)=\{0011,0101,0110,\\ 1001,1010,1100,1111\}=  \{46,47, 48, 49,50,51,52,53\}$, $L(\tau)=\{00,11\} =\{54,55\}$, $L(\omega)=\{00,11\} \\ =\{56,57\}$. 

There is an edge from vertex $a_1a_2 \dots a_p$ in $L(u)$ to vertex $a'_1a'_2 \dots a'_q$ in $L(v)$ 
when $uv$ is an edge of $G$ and $a_i=a'_j$, $i \in [1,p], j \in [1,q]$ ( here $a_i,a'_j$ represent the same variable in both equations $u,v$ and have the same value 0 or 1).  
Note that $a_1a_2 \dots a_p$ is a binary sequence  corresponding to the variables in equation $u$, and 
$a'_1a'_2 \dots a'_q$ is a binary sequence representing the variables in equation $v$.

\begin{figure}
  \begin{center}
   \includegraphics[scale=0.68]{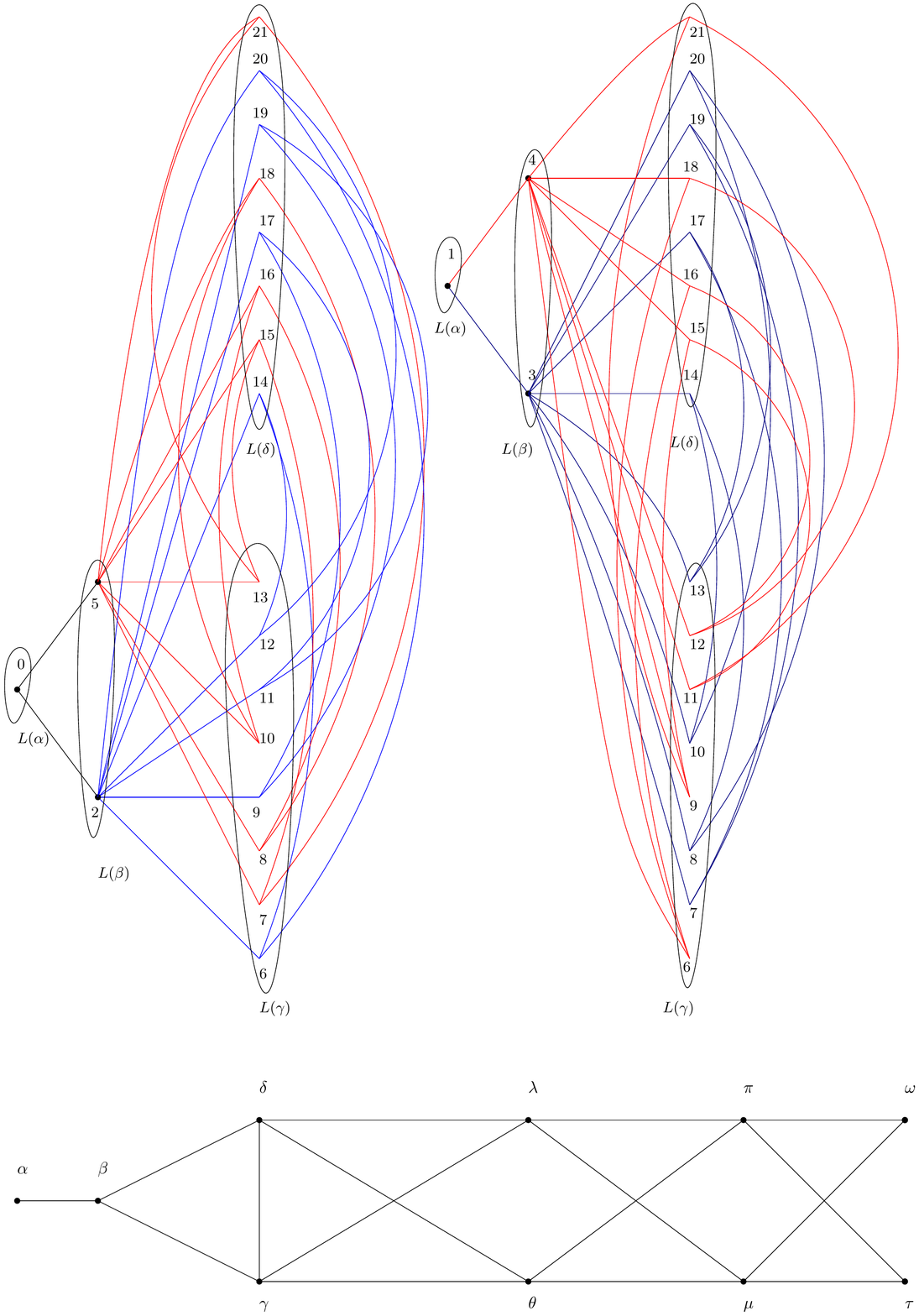}

  \end{center}
  \caption{ \small{The graph shows the list of vertices in $L(\alpha),L(\beta),L(\gamma),L(\delta)$, and their adjacency. The left figure shows $L_{\alpha,0}$  and the right figure shows $L_{\alpha,1}$ } }
\label{fig:G-12-1}
 \end{figure}

For example, there is an edge from $1100$ in $L(\lambda)$ to vertices $1111,1001,0101,0011$ in $L(\delta)$; the last bit in $L(\delta)$ corresponds to variable $X_{\delta \lambda}$ and the second bit in $L(\lambda)$ corresponds to variable $X_{\delta \lambda}$ and they both are $1$ (see Figure \ref{fig:G-12-1} ).  
One can define a Maltsev polymorphism $h(x;a,b,c)=a+b+c$ (sum is applied coordinate wise in $Z_2$). 
Suppose the Algorithm  \ref{alg-remove-minority-new} calls 
\Call{Sym-Dif} {$G,H,L,x=\alpha, a=1, b=0$}. Initially $G^{L,x}_{a,b}$ includes the vertices $\{\alpha,\beta,\gamma,\delta\}$ from $G$. The boundary set of $G'=G^{L,x}_{a,b}$ is $\{\gamma,\delta\}$. 
Looking at vertex $\gamma$ we would have :  \\
$L_{\alpha,1}(\gamma,\delta) =\{(6,16),(6,18),(9,15),(9,21),(11,15),(11,21),(12,16),(12,18),
(7,17),(7,19),(8,14),\\ (8,20), (10,14),(10,20),(13,17),(13,19)\}.$

$L_{\alpha,0}(\gamma,\delta) =\{(6,14),(6,20),(9,17),(9,19),(11,17),(11,19),(12,14),(12,20),(7,15),(7,21),(8,16),\\(8,18),(10,16),(10,18),(13,15),(13,21)\}$ (see Figure \ref{fig:G-12-1} right). 

\begin{figure}[ht]
  \begin{center}
   \includegraphics[scale=0.55]{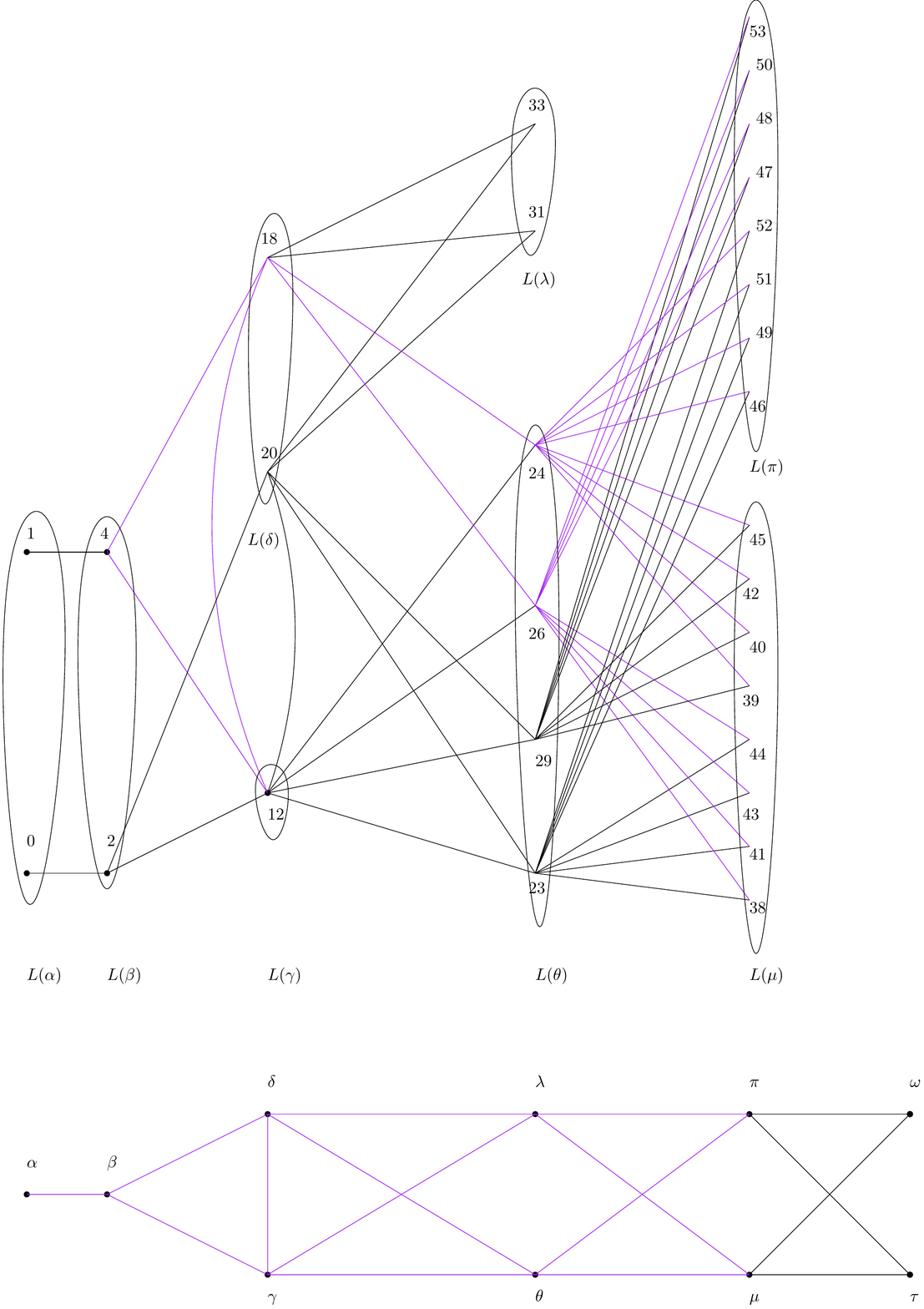}

  \end{center}
  \caption{ \small{Setting $L'(\gamma)=\{12\}$, $\delta$ and $18,20 \in L'(\delta)$ as a witness }}
\label{fig:G-12-6}
 \end{figure}

Let $L'=L_{\alpha,1}$. Since $L_{\alpha,1} (\gamma,\delta) \ne L_{\alpha,0}(\gamma,\delta)$, 
the \Call{Sym-Dif}{} continues adding the vertices of say $G^{L',\delta}_{20,18}$ into instance $G'$. Here according to the algorithm, $\delta$ with $20,18$ would witness $\alpha \in G$, and $1,0 \in L(\alpha)$ (see Figure \ref{fig:G-12-6}). Notice that in this situation $(12,18) \in L'(\gamma,\delta)$ but $(12,18) \not\in L_{\alpha,0}(\gamma,\delta)$ (as depicted in Figure \ref{fig:G-12-6}).  This means that $G'$ would include the vertices $\theta,\lambda,\mu,\pi$. Let $L_1$ be the restriction of $L'$ when $L'(\gamma)=\{12\}$, i.e. $L_1=L'_{\gamma,12}$.

\newpage

\begin{figure}[ht]
  \begin{center}
   \includegraphics[scale=0.52]{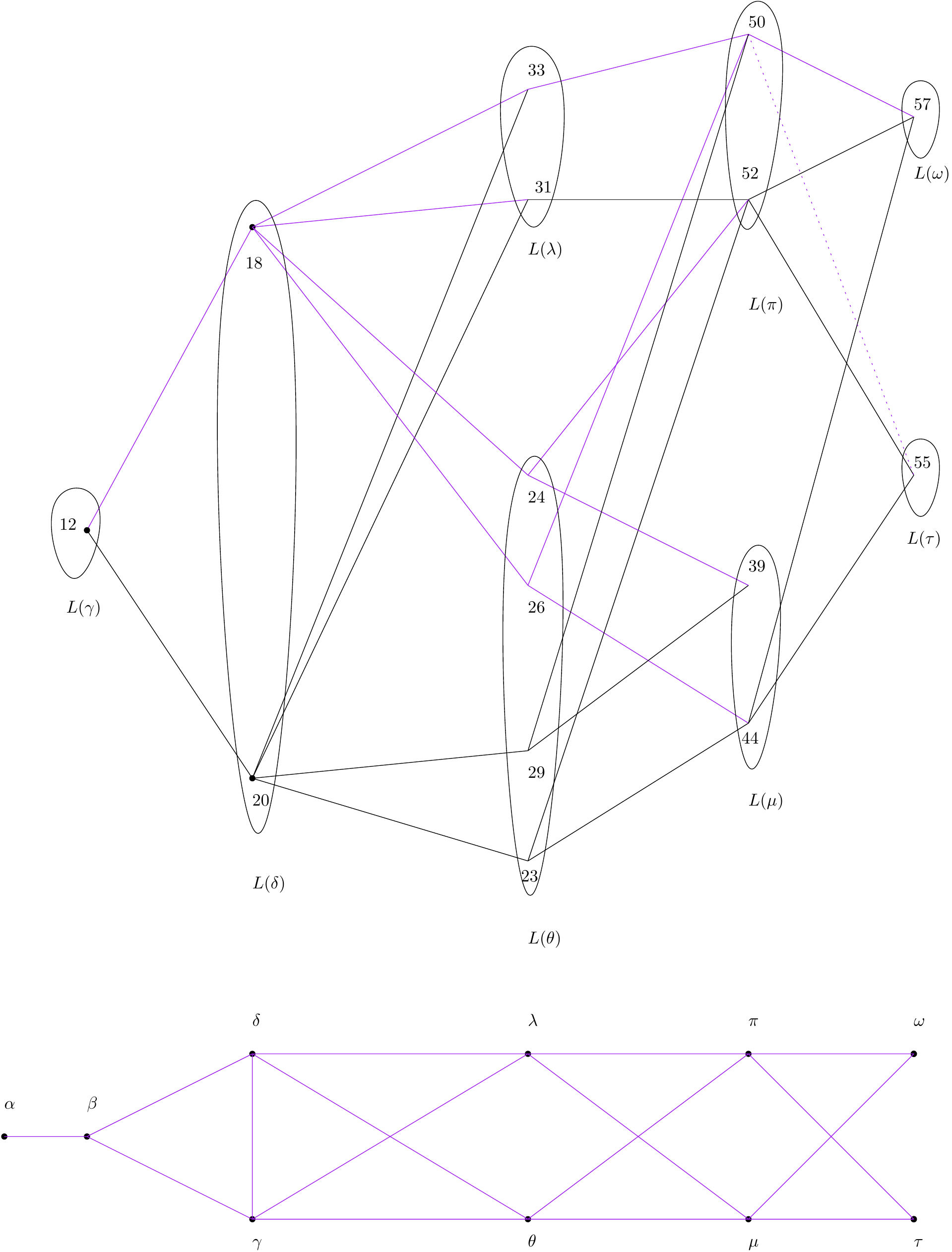}

  \end{center}
  \caption{$52,50 \in L_1(\pi)$ as a witness for $20,18 \in L_1(\delta)$ }
\label{fig:G-12-7}
 \end{figure}

Now vertex $\mu$ with $44 \in L_1(\mu)$ and vertex $\pi$ with two vertices $52,50 \in L_1(\pi)$ witness $\delta$ with $20,18 \in L_1(\delta)$. Thus we add $\tau,\omega$ into $G'$ (Figure \ref{fig:G-12-7} ).

Therefore, $G'$ includes all the vertices of $G$. It is not difficult to see that there is no solution for $G',L'$, and hence eventually $1$ is going to be removed from $L(\alpha)$.  

\newpage

\section{Proofs and Analysis}
In this section we prove the main result, Theorem \ref{thm:main}.  The proof is by induction on $\sum_{x \in V(G)}|L(x)|$. A key part of the algorithm is constructing sub-instances of the problem (on line \ref{line5} of the algorithm) and using a correct solution for these sub-instances to prune the lists  ( lines \ref{line7}, \ref{line8} of the algorithm).  Claim \ref{clm:algorithm-base-case} is for the base case of the induction. The main lemma needed to prove the algorithm correct is the following, which will be proved building on a number of claims.  
\begin{lemma}
\label{Maltsev-correctness}
Let $G, H, L$ be an instance of the list homomorphism problem just after line \ref{line8} of Algorithm \ref{alg-remove-minority-new}, i.e. after removing either $a,b$ from $L(x)$. 
If there is an $L$-homomorphism $g$ from $G$ to $H$ with $g(x) \in \{a,b\}$
before line \ref{line6}, then there is an $L$-homomorphism from $G$ to $H$ after removing $a$ or $b$ from $L(x)$.
\end{lemma}
\pf The proof if based on the induction on $\sum_{x \in V(G)} |L(x)|$. The base case for the Algorithm \ref{alg-remove-minority-new} is correct according to following claim.  

\begin{claim} \label{clm:algorithm-base-case}
Algorithm \ref{alg-remove-minority-new} is correct for the following base cases:  $|L(x)|=1$ for all $x\in V(G)$.
\end{claim}

Next we show that if we remove $a$ from $L(x)$ in line \ref{line7} then there is no homomorphism from $G$ to $H$ that maps $x$ to $a$. 

\begin{claim}\label{claim1}
Let $G, H, L$ be an instance of the list homomorphism problem just before line \ref{line7} of Algorithm \ref{alg-remove-minority-new}.  If $g^x_{a,b}$  does not exist, then there is no homomorphism from $G$ to $H$ that maps $x$ to $a$. 
\end{claim}
\noindent \textbf{Proof of Claim \ref{claim1}.}
By induction hypothesis, we 
 may assume that the algorithm returns a correct answer for the smaller instance $G',L',H$, provided that this instance has Maltsev polymorphism with respect to the lists $L'$. Note that $G'=\widehat{G^{L,x}_{a,b}}$ returned from \textsc{Sym-dif} is an induced sub-digraph of $G$, and that for each $y \in G'$, $L'(y)$ contains all $c\in L(y)$ such that $(a,c)\in L(x,y)$. Thus, it is easy to see that $G',L',H$ admits a Maltsev polymorphism with respect to the lists.

Suppose there were an $L$-homomorphism $g$ from $G$ to $H$ with $g(x)=a$ . For each $y \in G'$, $(a,g(y))\in L(x,y)$ and also $(a,g(y))\in L'(x,y)$.
Observe that if there exists such a $g$, then $g$ would lie within the $L'$ lists. 
If RemoveMinority is correct on $G', H, L'$ then the claim is proved. 
Note that $\sum_{y \in V(G')}|L'(y)| < \sum_{y \in V(G)}|L(y)|$. 
By Claim \ref{clm:algorithm-base-case}, RemoveMinority is correct for $G$ with a singleton list vertex, and thus the claim is proved by induction. 
\qed 
\\

\noindent{ \textbf{Remark :}} According to the algorithm, in performing the steps on a smaller instance we don't deal with vertex $x$ anymore. Thus, removing an element from $L(x)$ in lines \ref{line7}, \ref{line8} would not affect the outcome of the sub-routine \Call{Sym-Dif}{}.

Suppose before removing $b$ from $L(x)$ on line \ref{line8} of Algorithm \ref{alg-remove-minority-new}, there exists an $L$-homomorphism $g$ from $G$ to $H$ with $g(x)=b$. We will show that after removing $b$, there remains an $L$-homomorphism $\psi$ from $G$ to $H$ with $\psi(x)=a$.  
The construction of $\psi$ is broken into two different cases based on the lists in $G^{L,x}_{a,b}$, and the image of $g$ over $G$, namely whether there exists a certain rectangle/sub-graph between the two or not. 
\paragraph{If no rectangle exists :}  This means there is no vertex $z \in \widehat{G^{L,x}_{a,b}}$ so that $(a,g(z)) \in L(x,z)$ (note that since $g$ is a homomorphism, $(b,g(z)) \in L(x,z)$).  Therefore, $\widehat{G^{L,x}_{a,b}}$ extends all over $G$. In this case since $g^x_{a,b}(y)$ exists, $g^x_{a,b}(y)$ is an $L$-homomorphism from $G$ to $H$ that maps $x$ to $a$.

\paragraph{There exists a rectangle:} 
The argument is based on the following claims, and  the structure of the proof is as follows. In order to prove the statement of the Lemma \ref{Maltsev-correctness} we use Claim \ref{cl1}. The proof of Claim \ref{cl1} is based on the induction on the size of the lists, and using the assumption that the statement of Lemma \ref{Maltsev-correctness} is correct for a smaller instance.  

\begin{claim}\label{cl2}
For every two vertices $z,z_1 \in B'=B(\widehat{G^{L,x}_{a,b}})$ defined in line \ref{line31} of the function \Call{Sym-Diff}{}, we have
$(g^{x}_{a,b}(z_1),g(z)) \in L_{x,a}(z_1,z)$. 
\end{claim} 

\begin{claim}\label{cl1}
Consider the lists $L$ after line \ref{line8}. Let $G_1=\widehat{G^{L,x_1}_{c_1,c_2}}$ when $g^{x_1}_{c_1,c_2}$ exists. Let $a_1,b_1 \in L_{x_1,c_1}(y)$, and suppose there exists an $L$-homomorphism $g^{y}_{a_1,b_1}$ from $\widehat{G^{L,y}_{a_1,b_1}}$ to $H$. Let $(G'_1,L'_1)=$ \Call{Sym-Dif} {$G_1,H,L_{x_1,c_1},y,a_1,b_1$}. 
Then \Call{RemoveMinority}{$G'_1,H,L'_1$} returns a non-empty homomorphism. In other words, if a small test passes for $(G,L,H)$ it also passes for $G_1,L_{x_1,c_1},H$. 
\end{claim}

\begin{claim}\label{cl3}
Suppose there exists an L-homomorphism $\phi$ from $G^{L,x}_{a,b} \setminus B(G^{L,x}_{a,b})$ to $H$ with $\phi(x)=a$. Suppose there exists an $L$-homomorphism $\psi$ from $G$ to $H$ with $\psi(x)=b$. Moreover, for every two vertices $y,z \in B(G^{L,x}_{a,b})$, $L_{x,a}(y,z) = L_{x,b}(y,z)$. Then there exists an $L$-homomorphism $\phi'$ from $G^{L,x}_{a,b}$ to $H$ such that $\phi'(x)=a$, and  $\phi'(y)=\psi(y)$ for every $y \in B(G^{L,x}_{a,b})$. 
\end{claim}

By Claim \ref{cl3} we may assume the partial homomorphism $g^x_{a,b}$ is extended to the boundary vertices of $B(\widehat{G^{L,x}_{a,b}})$. This is because $\widehat{G^{L,x}_{a,b}}$ is constructed by gluing some $G^{y,L}_{c_1,c_2}$.  

\paragraph{Collapsing the list on the boundaries:}
Suppose Claim \ref{cl1} holds. Now we let $G_1 =\widehat{G^{L,x}_{a,b}}$,  $x_1=x$, and $c_1=a$ in Claim \ref{cl1}. If $G \setminus \widehat{G^{L,x}_{a,b}} = \emptyset$ then $g^x_{a,b}$ is a homomorphism from $G$ to $H$ and we are done. Otherwise, let $z$ be a vertex in $B(\widehat{G^{L,x}_{a,b}})$. We may assume $z$ is chosen such that $g(z)=c \ne g^x_{a,b}(z)=d$. If there is no such $z$ then we define $\phi(y)=g^x_{a,b}(y)$ for every $y \in \widehat{G^{L,x}_{a,b}}$, and $\phi(y)=g(y)$ for every $y \in G \setminus \widehat{G^{L,x}_{a,b}}$. It is easy to see that $\phi$ is a L-homomorphism from $G$ to $H$ with $\phi(x)=a$. 
Thus, we proceed by assuming the existence of such a $z$. 
Since $G,L_{x,a}$ is smaller than the original instance (and the small test passes for $G,L_{x,a}$ by Claim \ref{cl1}), we may assume that if there exists an $L_{x,a}$-homomorphism from $\widehat{G^{L,x}_{a,b}}$ to $H$ that maps $z$ to $d$ (and $x$ to $a$) then there exists an $L_{x,a}$-homomorphism $g_0 : \widehat{G^{L,x}_{a,b}} \rightarrow H$ that maps $z$ to $c$. 
Thus, we may reduce the lists $L_{x,a}$ by collapsing the lists in $\widehat{G^{L,x}_{a,b}}$ by identifying $c$ and $d$ in $L_{x,a}(z)$ into $c$. This would mean we remove $d$ from $L_{x,a}(z)$. Notice that by removing $d$, and applying Preprocessing, the remaining lists which are $L_{x,a,z,c}$, have the property that the small tests pass inside them. This follows from Claim \ref{cl1}, because we are restricting the lists by setting $L(x)=\{a\}$, and $L(z)=\{c\}$. \\

\noindent {\textbf{The key observation:}} We need to observe that according to the construction of $\widehat{G^{L,x}_{a,b}}$, for every other vertex $z_1 \in B(\widehat{G^{L,x}_{a,b}})$, we have $(c,g(z_1)) \in L_{x,a}(z,z_1)$. This follows from Claim \ref{cl2}.



Next, consider vertex $z_1 \ne z$ from 
$B(\widehat{G^{L,x}_{a,b} })$. If for every $z_1 \in B(\widehat{G^{L,x}_{a,b}}) \setminus \{z\}$, 
$g(z_1)=g^x_{a,b}(z_1)$  then we define the desired homomorphism $\psi : G \rightarrow H$, as follows, $\psi(y)=g_{0}(y)$ for every $y \in \widehat{G^{L,x}_{a,b}}$ and $\psi(y)=g(y)$ for every $y \in G \setminus G^{L,x}_{a,b}$.
Thus, we continue by assuming $c_1=g(z_1) \ne d_1=g^x_{a,b}(z_1)$.
By the observation, $(c,c_1) \in L_{x,a}(z,z_1)$. Consider the lists $L_1=L_{x,a,z,c}$. By Claim \ref{cl1}, since there exists an $L_1$-homomorphism, $g_0$ from $\widehat{G^{L,x}_{a,b}}$ to $H$, then there exists an $L_1$-homomorphism $g_1$ that maps $x$ to $a$ and $z$ to $c$ and $z_1$ to $c_1$. This would mean we remove $d_1$ from $L_{1}$ lists. Therefore, $g_1$ agrees with $g$ on $z,z_1$. By this procedure, we continue collapsing the rest of the list of the vertices on the boundary of $ \widehat{G^{L,x}_{a,b}}$ accordingly. At the end, if there is no boundary vertex left then as we argued, there exists a homomorphism 
$\psi$ from $G$ to $H$ with $\psi(x)=a$. \\



\noindent{\textbf {Proof of Claim \ref{cl2}.}}
We use induction on the size of $G'= \widehat{G^{L,x}_{a,b}}$. If $B'=B(G^{L,x}_{a,b})$ then by definition of $B(G^{L,x}_{a,b})$, for every $z,z_1 \in B'$, $(c',d') \in L_{x,a}(z,z_1)$ iff $(c',d') \in L_{x,b}(z,z_1)$ and the claim is proved. Otherwise, there must exist two vertices $u,v$ and $c_1,c_2 \in L(u)$ and $d_1,d_2 \in L(v)$ so that $u,c_1,c_2$ would witness $x,a,b$ at $v,d_1,d_2$. This would imply that   $(c_1,d_1),(c_2,d_2) \in L_{x,a}(u,v) \setminus L_{x,b}(u,v)$, and $(c_1,d_2),(c_2,d_1) \in L_{x,b}(u,v) \setminus L_{x,a}(u,v)$. Let $L_1=L_{x,a}$. We may assume that $g(u)=c_2$. Let $c_1=g^{x}_{a,b}(u)$. 
According to the algorithm, if both $z_1,z$ belong to $(\widehat{G^{L_1,u}_{c_1,c_2}} \cap G')  \setminus (G^{L,x}_{a,b} \setminus B(G^{L,x}_{a,b}))$ then by induction hypothesis, starting at $u$ and $c_1,c_2 \in L_1(u)$, 
we conclude that  $(g^{x}_{a,b}(z_1),g(z)) \in L_1(z_1,z)$. Similarly if both $z_1,z$ belong to $(G' \cap \widehat{G^{L_1,v}_{c_1,c_2}}) \setminus (G^{L,x}_{a,b} \setminus B(G^{L,x}_{a,b}))$ then $(g^{x}_{a,b}(z_1),g(z)) \in L_1(z_1,z)$. 
If $z_1 \in (\widehat{G^{L_1,u}_{c_1,c_2}} \cap G') \setminus (G^{L,x}_{a,b} \setminus B(G^{L,x}_{a,b}))$ and $z \in (\widehat{G^{L_1,v}_{c_1,c_2}} \cap G') \setminus (G^{L,x}_{a,b} \setminus B(G^{L,x}_{a,b})) $ then because $c_1,c_2 \in L_{x,a}(u)$ and $d_1,d_2 \in L_{x,a}(v)$, we have, $(g^{x}_{a,b}(z_1),g(z)) \in L_1(z_1,z)$. \qed \\

\noindent{\textbf {Proof of Claim \ref{cl1}.}} Note that there exists an $L$-homomorphism $g^y_{a_1,b_1}$ from $\widehat{G^y_{a_1,b_1}}$ to $H$. Let $L_1=L_{x_1,c_1}$. 
Now for every $z_1 \in  G_1$, let 
$L_2(z_1)= \{g^{y}_{a_1,b_1}(z_1) \}$ if $g^{y}_{a_1,b_1}(z_1) \in L_1$. 
Otherwise, set $L_2(z_1)= \{ a' | a' \in L_{1}(z_1), (a',a_1) \in L_1(z_1,y)\}  \cup \{ g^{y}_{a_1,b_1}(z_1) \}$. 
Note that $L_2(y)=\{a_1\}$. If $L_1=L_2$ then we are done. Otherwise, suppose $L_1 \subset L_2$. {\em We use induction  $\sum_{y \in G_1} |L_1(y)|$.}

Notice that $\widehat{G^{L,y}_{a_1,b_1}} \cap G_1 = \widehat{G^{L_1,y}_{a_1,b_1}} \cap G_1$. This is because both of these instances are constructed based on the existence of the rectangles. So $(L_1)_{y,a_1}(u) \ne \emptyset$ for $u \in \widehat{G^{L_1,y}_{a_1,b_1}} \cap G_1$.

\begin{figure}[ht]
  \begin{center}
   \includegraphics[scale=0.5]{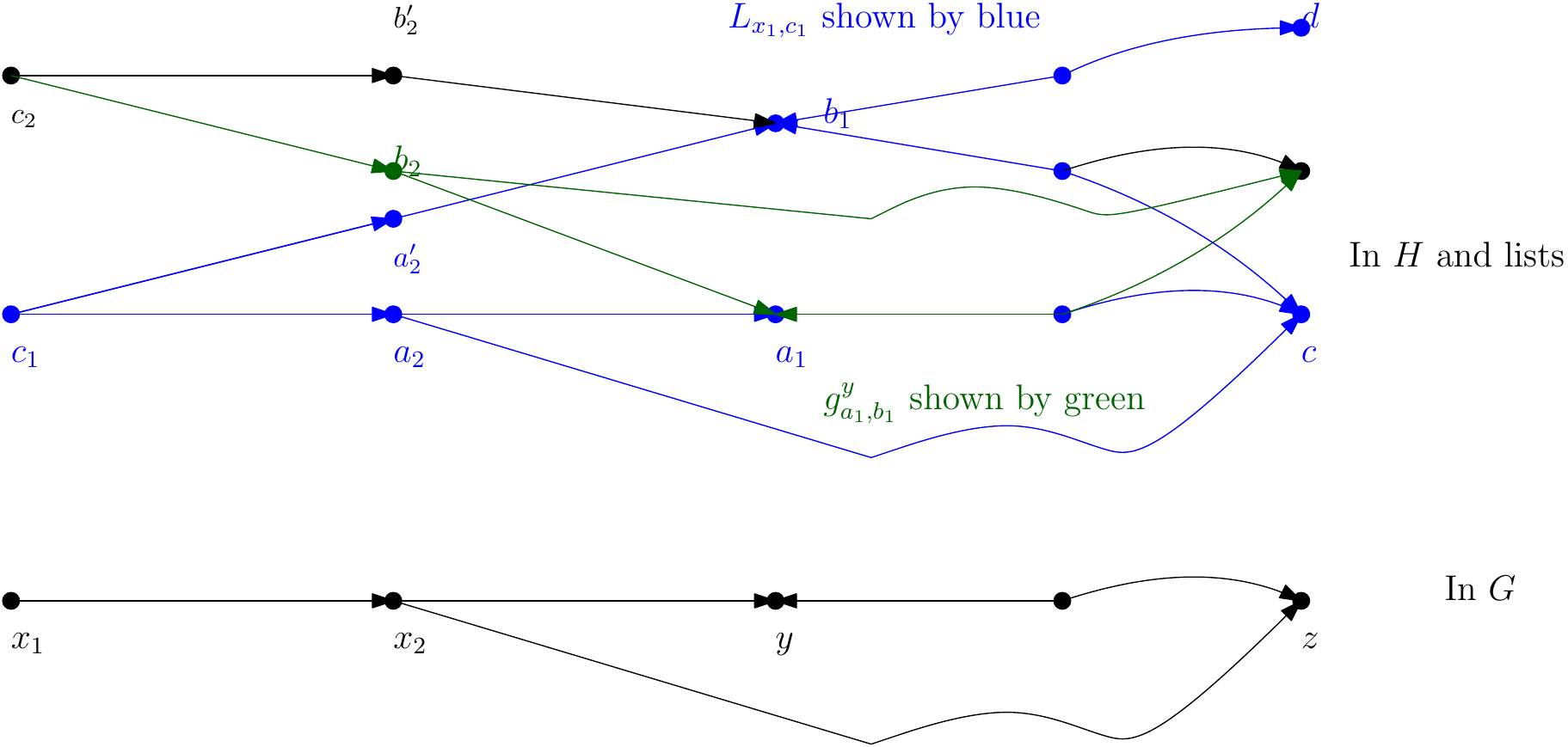}
  \end{center}
  \caption{ \small{Illustration that the small test passes in $L_{x_1,c_1}$ }}
\label{fig:4-9-1}
 \end{figure}

\noindent Let $Y$ be an oriented path from $x_1$ to $y$, and let $x_2$ be a vertex on $Y$ so that $b_2=g^{y}_{a_1,b_1} (x_2) \not \in L_1(x_2)$. We may assume $c_2=g^y_{a_1,b_1}(x_1)$. 
By definition $x_1,c_1,c_2$ and $y,a_1,b_1$ induce a rectangle in $Y$ and $L(Y)$. Let $a_2 \in L_1(x_2)$, and note that by definition $a_2$ exists, because $(c_1,a_1) \in L_1(x_1,y)$ (Figure \ref{fig:4-9-1}).

\begin{figure}[ht]
  \begin{center}
   \includegraphics[scale=0.6]{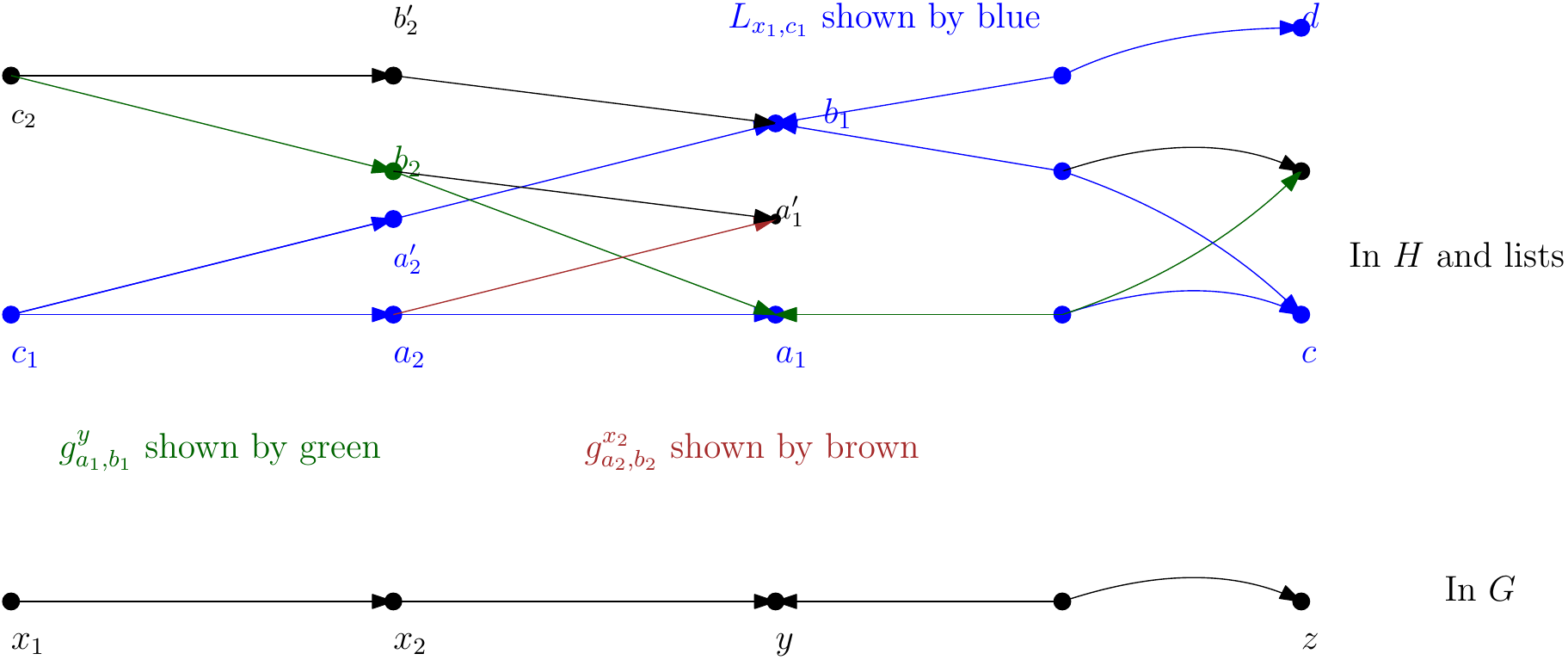}
  \end{center}
  \caption{ Second Illustration that the small test passes in $L_{x_1,c_1}$ }
\label{fig:4-9-2}
 \end{figure}

By assumption, $g^{x_2}_{a_2,b_2}$ exists. First suppose $g^{x_2}_{a_2,b_2}(x_1)=c_1$. Let $a'_1=g^{x_2}_{a_2,b_2}(y)$ (note that $a'_1 \in L_1(y)$ because $g^{x_2}_{a_2,b_2}(x_1)=c_1$). Now consider the lists $L'_2=(L_1)_{x_2,a_2}$. Suppose $a'_1 \ne a_1$ (see Figure \ref{fig:4-9-2}).  Observe that $g^{y}_{a_1,a'_1}$ exists (otherwise $a_1$ wouldn't be in $L(y)$). By induction hypothesis, for instance $G_2= \widehat{G^{L,x_2}_{a_2,b_2}} \cap G_1$ and $L'_2$ (notice that $L'_2 \subset L_1$; i.e. the total elements inside the lists $L'_2$ is strictly smaller than the total elements of $L_1$, and $G_2$ is a subset of $G_1$), all the small tests pass.


Notice that for the instance $G_2,L'_2$,  by applying the collapsing argument (on the $L'_2$ lists ), we conclude, there exists $L'_2$-homomorphism, $g_2$ from $G_2$ to $H$ such that $g_2(y)=a_1$, and hence, 
$g_2(Y)$ lies in $L_1$. By applying this argument, we consider other vertices of $G_2$, say $z_1$ for which $g_2(z_1) \not\in L_{1}(z_1)$, and by induction hypothesis for $L'_2$, we further modify $g_2$ so its image lies in $L_1$.


Next suppose $c'_1=g^{x_2}_{a_2,b_2}(x_1) \ne c_1$. We may assume $x_2$ is in such a way that there is no other rectangle in $Y,L(Y)$, involving $x_1,x_2$  with $c_1,c'_1 \in L(x_1)$ and  $a_2,a'_2 \in L(x_2)$ ($a_2 \ne a'_2$). Notice that this can be done by just taking the last rectangle containing $x_1,c_1$ on $Y$ and $L(Y)$. Notice that since $c_1 \in L(x_1)$, the homomorphism $g^{x_1}_{c_1,c'_1}$ exists. By the assumption about $x_2$, it is easy to see that $g^{x_1}_{c_1,c'_1}(x_2)=a_2$. Let $g_2=g^{x_1}_{c_1,c'_1}$ and consider the lists $L'_2=L_{x_1,c_1,x_2,a_2}$. Now as in the previous case we continue considering other vertices of $G_1$, say, $z_1$ for which $g^{y}_{a_1,b_1}(z_1) \not\in L_{x_1,c_1}(z_1)$, and by induction hypothesis for $L'_2$ we would further alter $g_2$ so that its image lies in $L_{1}$.

Since the choice of $x_2$ was arbitrary, we conclude that the small tests on instance \Call{Sym-Dif}{$G_1,L_1,y,a_1,b_1$} would pass. Thus, we may again use the collapsing argument, and assume that $g^y_{a_1,b_1}$ maps $x_1$ to $c_1$, and by doing so, as we argued for $g_2$, we continue modifying $g^y_{a_1,b_1}$ until its images lies in $L_1$. \\

\noindent{\textbf {Proof of Claim \ref{cl3}.}}
Let $P$ be a path from $x$ to $y \in B(G^{L,x}_{a,b})$ and $Q$ be a path from  $x$ to $z \in B(G^{L,x}_{a,b})$. 
Suppose $v$ is the intersection vertex of $P$ and $Q$ ($v$ could be $x$). Let $c_1,c_2 \in L(y)$ such that $x,a,b$, and $y,c_1,c_2$ induce a rectangle. Let $d_1,d_2 \in L(z)$ such that $x,a,b$ and $z,d_1,d_2$ induce a rectangle. 

\begin{figure}
  \begin{center}
   \includegraphics[scale=0.6]{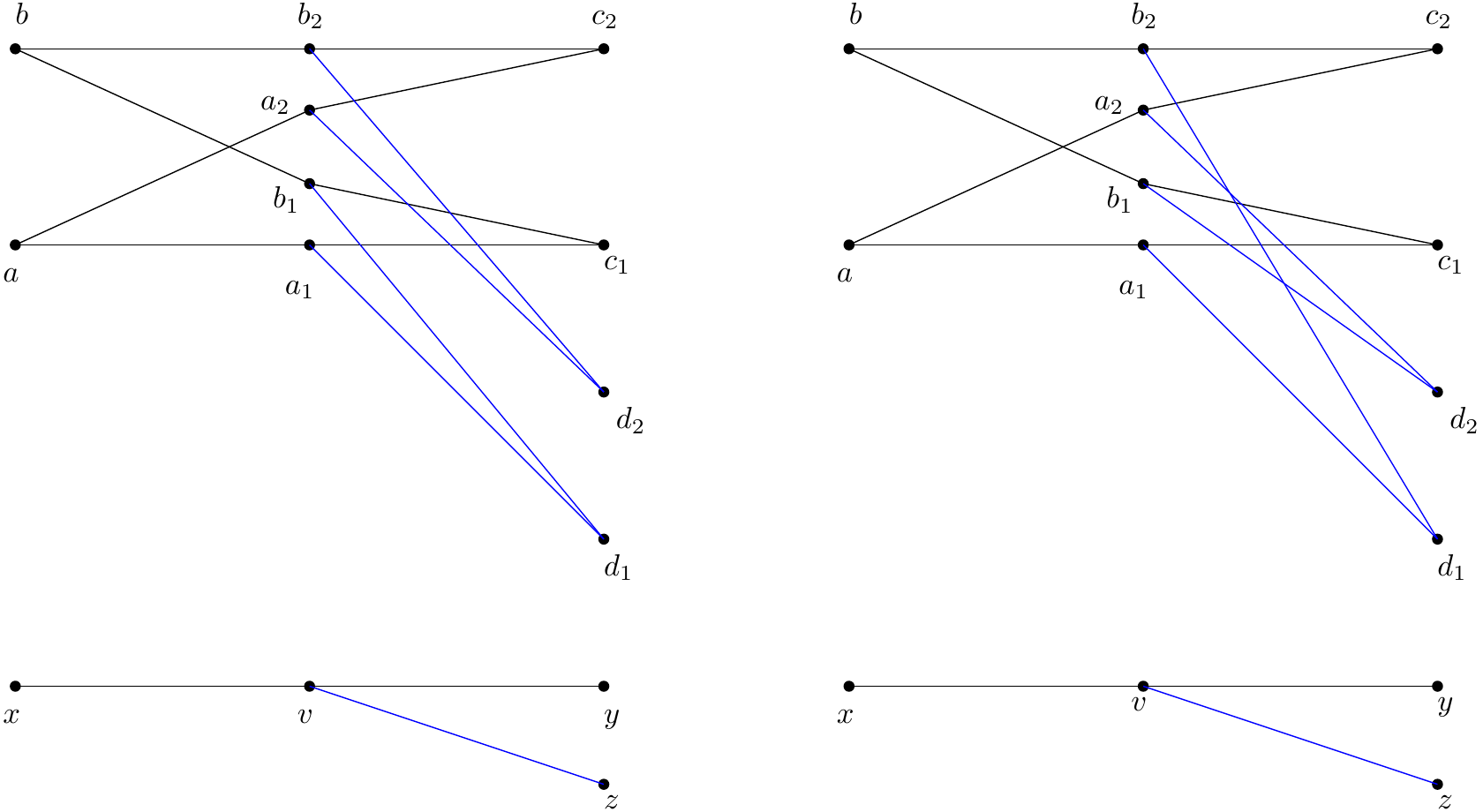}
  \end{center}
  \caption{ \small{On the left figure the pair list for boundary vertices $y,z$ are the same while on the right figure the pair lists for $y,z$ are disjoint } }
\label{fig:list-get-smaller}
 \end{figure}

Consider $a_1,a_2 \in L_{x,a}(v)$ and two vertices $c_1,c_2 \in L_{x,a}(y)$ such that $(a,a_1),(a,a_2) \in L_{x,a}(v)$ and $(a_1,c_1),(a_2,c_2) \in L_{x,a}(v,y)$. Let $d_1,d_2 \in L_{x,a}(z)$ such that $(a_1,d_1),(a_2,d_2) \in L_{x,a}(v,z)$. Since $L_{x,a}(y,z) = L_{x,b}(y,z)$, we may assume that $(b_1,d_1),(b_2,d_2) \in L_{x,b}(y,z)$ (see Figure \ref{fig:list-get-smaller} left). We may assume $\psi(y)=c_2$, and $\psi(z)=d_2$. Now we may assume there exists an $L$-homomorphism $\phi_1 : G^{L,x}_{a,b} \rightarrow H$ with $\phi_1(x)=a$ and $\phi_1(v)=a_2$. This is by induction on the total sizes of the lists, $L_{x,a}$ and using Claim \ref{cl2} that the small test pass inside $L_{x,a}$. Let $y_1y \in P$ be the last arc (forward or backward). 
Now by induction hypothesis for $L_{x,a,v,a_2}$ we can further modify $\phi_1$ such that $\phi_1(y_1) \psi(y)$ is an arc of $H$ (with the same direction as $y_1y$). Similarly we can modify $\phi_1$ so that for the last arc (forward or backward) of $Q$ say $z_1z$, $\phi_1(z_1)\phi(z)$ is an arc of $H$ (with the same direction as $z_1z$). We continue modifying $\phi_1$ until for every path $W$ from $x$ to $w \in B(G^{L,x}_{a,b})$ the last arc (forward or backward) of $W$ say $w_1w$, we have $\phi_1(w_1)\psi(w)$ is an arc of $H$ (with the same direction of $w_1w$). Now we set $\phi'(u)=\phi_1(u)$ for every $u \in G^{L,x}_{a,b} \setminus B(G^{L,x}_{a,b})$ and for every $u \in B(G^{L,x}_{a,b})$ set $\phi'(u)=\psi(u)$. 

 \qed




\begin{lemma}
The Algorithm \ref{alg-remove-minority-new} runs in polynomial time. More precisely, the running time of the algorithm is $\mathcal{O}(|G|^4|H|^5)$. 
\end{lemma}
\pf According to our assumption we removed the twin vertices $a,b \in L(x)$. Thus, the clique instances, where $G$ is a complete graph and for every $a \in L(x)$ is adjacent to every vertex in $L(y)$, $y \in V(G)$ are just simply out of questions. When the lists are singletons, it takes $O(|G|)$ to verify whether there is a homomorphism from $G$ to $H$.
Moreover, we handle each connected component of $G \times_L H$ independently. This means the running time of the instance is the sum of the running time 
of each connected component. So if $T(G,L)$ is the running time of the instance $G,L,H$ then we can express $T(G,L)$ as $\sum_{i=1}^t T(G,L_i)$ where 
$L= \cup_{i=1}^r L_i$, and  $L_i$'s are disjoint. 

Suppose $\widehat{G^{L,x}_{a,b}}=G^{L,x}_{a,b} \setminus B(G^{L,x}_{a,b})$. Now consider the instances
$(G_1,L_1)=$ \Call{Sym-Dif}{$G,H,L,x,a,b$} and 
$(G_2,L_2)=$ \Call{Sym-Dif}{$G,H,L,x,b,a$}. Notices that $G_1=G_2$, but $L_1,L_2$ are disjoint. If during the recursion step we end up with sub-instances in which the pair lists on the boundary are the same then we further split the lists. Therefore, the running time would be $\mathcal{O}(poly_1(|G|)poly_2(|H|))$. The following claim would give a more structural property of the  pair lists on the boundary vertices. We use the existence of the Maltsev polymorphism yet again. 

\begin{claim}\label{partition}
For every $y,z \in B(G^{L,x}_{a,b})$, and every two distinct elements $c_1,c_2 \in L_{x,a}(y) \cap L_{x,b}(y)$, exactly one of the following occurs. 
\begin{itemize}
    \item [1.] $\{ d_1,d_2  | \ \ (c_1,d_1),(c_2,d_2) \in L_{x,a}(y,z) \}  = \{ d_1,d_2  | \ \ (c_1,d_1),(c_2,d_2) \in L_{x,b}(y,z) \} $
    
    \item [2.]  $\{ d_1,d_2  | \ \ (c_1,d_1), (c_2,d_2) \in L_{x,a}(y,z) \}  \cap 
    \{  d_1,d_2  | \ \ (c_1,d_1),(c_2,d_2) \in L_{x,b}(y,z) \}= \emptyset$.
    
\end{itemize}
\end{claim}
\pf Let $P$ be an arbitrary oriented path from $x$ to $y$ in $G^{L,x}_{a,b}$.  By definition of $B(G^{L,x}_{a,b})$, $L_{x,a}(y)=L_{x,b}(y)$. 
Let $Q$ be an arbitrary oriented path from $x$ to $z$ in $G^{L,x}_{a,b}$. 
Let $v$ be a vertex in the intersection of $P,Q$. Since $(a,c_1),(a,c_2) \in L_{x,a}(x,y)$, there exist $a_1,a_2 \in L_{x,a}(v)$ such that $(a_1,c_1),(a_2,c_2) \in L_{x,a}(v,y)$. 

Let $b_1 \in L_{x,b}(v)$ so that $(b_1,c_1) \in L_{x,b}(v,y)$. 
Let $b_2 \in v$ such that $h(v;a_1,a_2,b_1)=b_2$. This means that 
$(b_2,c_2) \in L_{x,b}(v,y)$ (note that when $x$ is the only intersection of $P,Q$ we have $a=a_1=a_2$, and $b=b_1=b_2$). 
Observe that $x,a,b,y,c_1,c_2$ induce a rectangle in $L(P)$ (see Figure \ref{fig:list-get-smaller}).

Let $N(a_1)=\{d  | (a_1,d) \in L_{x,a}(v,z) \}$, $N(a_2)=\{ d | (a_2,d) \in L_{x,a}(v,z) \}$, and 
let $N(b_1)=\{d  | (b_1,d) \in L_{x,b}(v,z) \}$, $N(b_2)=\{ d  | (b_2,d) \in L_{x,b}(v,z) \}$. By the rectangle property, 
either $N(c') \cap N(c'') =\emptyset$ or $N(c')=N(c'')$ for every $c',c'' \in \{a_1,a_2,b_1,b_2 \}$. Suppose $d_1 \in N(a_1)$, $d_2 \in N(a_2)$. Thus $
(c_1,d_1),(c_2,d_2) \in L_{x,a}(y,z)$. 

\noindent Up to symmetry, there are three possibilities : 

\begin{itemize}
    \item[a)]  $d_1 \in N(b_1)$ or $d_2 \in N(b_2)$ 
    \item[b)] $d_1 \in N(b_2)$ or $d_2 \in N(b_1)$ 
    \item [c)] $(N(a_1) \cup N(a_2)) \cap (N(b_1) \cup N(b_2)) =\emptyset$ 
\end{itemize}

\noindent \textbf{ Case a.}  $d_1 \in N(b_1)$ or $d_2 \in N(b_2)$.  Suppose the first situation occurs. First suppose $d_1 \in N(b_1)$. Now by applying the polymorphism definition on the oriented path in $L_{x,a} \cup L_{x,b} ( Q)$ we  conclude that there is an oriented path from $b_2$ to $d_2$ in $L_{x,b}$. This  implies that $(d_1,c_1),(d_2,c_2) \in L_{x,a}(y,z)$ and $(d_1,c_1),(d_2,c_2) \in L_{x,b}(y,z)$. 

Now suppose $d_2 \in N(b_2)$. Let 
$b'_1 =h(y,a_1,a_2,b_2)$. By applying the polymorphism definition on the oriented path in $L_{x,a} \cup L_{x,b} (P)$, we observe that $(b,b'_1)\in L_{x,b}(x,y)$ and $(b'_1,c_1) \in L_{x,b}(y,u)$. 
Now by applying the polymorphism definition 
on the oriented path in $L_{x,a} \cup L_{x,b} (Q)$, we conclude that there is an oriented path from $b'_1$ to $d_1$ in $L_{x,b}$. This implies that $(d_1,c_1),(d_2,c_2) \in L_{x,a}(y,z)$ and $(d_1,c_1),(d_2,c_2) \in L_{x,b}(y,z)$. 

Finally, when $N(a_1)=N(b_2)$, then by applying the  polymorphism definition on the path in $L_{x,a} \cup L_{x,b} (Q)$, we observe that $(b'_1,d_1), (b'_1,d_2),(b_2,d_1),(b_2,d_2) \in L_{x,b}(v,z)$. Therefore,  
$(c_1,d_1),(c_1,d_2),\\(c_2,d_1),(c_2,d_2) \in L_{x,a}(y,z) $ as well as $(c_1,d_1),(c_1,d_2),(c_2,d_1),(c_2,d_2) \in L_{x,b}(y,z)$, and hence, 
 $\{ d_1 \in L_{x,a}(z) \mid (c_1,d_1) \in L_{x,a}(y,z) \}  =
    \{ d_2 \in L_{x,b}(z) \mid (c_1,d_2) \in L_{x,b}(y,z) \}$. This proves (1). \\


\noindent \textbf{ Case b.}  $d_1 \in N(b_1)$ or $d_2 \in N(b_2)$. We may assume that $N(a_1) \cap N(a_2)= \emptyset$, as otherwise, situation in Case 1 occurs. Analogously, we assume that $N(b_1) \cap N(b_2) =\emptyset$, as otherwise, situation $1$ occurs. In this case $(c_1,d_1),(c_2,d_2) \in L_{x,a}(y,z)$, and $(c_1,d_2),(c_2,d_1) \in L_{x,b}(y,z)$. Thus, 
 $\{ d_1,d_2 \in L_{x,a}(z) \mid (c_1,d_1),(c_2,d_2) \in L_{x,a}(y,z) \}  \cap \{ d_1,d_2 \in L_{x,b}(z) \mid (c_1,d_1),(c_2,d_2) \in L_{x,a}(y,z) \} = \emptyset$. This proves (2). \\

\noindent \textbf{ Case c.} Since   $(N(a_1) \cup N(a_2))$ and  $(N(b_1) \cup N(b_2))$ have no intersection, none of the situations in Case 1 occurs, and hence, (2) holds. \qed \\

\noindent By Claim \ref{partition}, the following hold. 
\begin{itemize}
    \item Suppose $y \in B(G^{L,x}_{a,b})$, and $c_1,c_2 \in L_{x,a}(y)$ witness $x,a,b$ in $z \in B(G^{L,x}_{a,b})$. Then $\{ d_1,d_2  \mid  (c_1,d_1), (c_2,d_2) \in L_{x,a}(y,z) \}  \cap 
    \{  d_1,d_2  | \ \ (c_1,d_1),(c_2,d_2) \in L_{x,b}(y,z) \}= \emptyset$.
\end{itemize}

\noindent Note that the lists in $L_{x,a}$ and $L_{x,b}$ are disjoint for every vertex $y \in G^{L,x}_{a,b} \setminus B(G^{L,x}_{a,b})$.


So we can say the instance of the problem contains the rectangles of $G \times_L H$, all containing $x,a,b$ or if it contains some other vertices beyond $B(G^{L,x}_{a,b})$ then the pair lists are partitioned in the vertices of $G^{L,x}_{a,b}$ . It follows from the above observation that in the instance ($\widehat{G^{L,x}_{a,b}},L_{x,a},H$), the lists $L_{x,a,u,c_1}(v)$, $L_{x,a,u,c_2}(v)$ are disjoint. This means in the next step of algorithm, we further end up with disjoint lists. 
Therefore, essentially we partition the instances into $|G|^2|L|^2$ sub-instances. 
If the running time of each instance $(G_1,L_1)$ is a polynomial of $poly_1(|G_1|)poly_2(|L_1|)$ then the running time of instance $G,L$ would be 
$\mathcal{O}( poly_1(|G|)poly_2(|L|))$.

We run the Preprocessing and it takes $\mathcal{O}(|G|^3|H|^3)$, and it takes $O(|G||H|^2)$ to construct each instance $\widehat{G^{L,x}_{a,b}}$. So the running time for sub-instance  $\widehat{G^{L,x}_{a,b}},L_{x,a}$ is $\mathcal{O}(|G|^3|H|^3)$. 
There are $|G||H|^2$ such instances, and because $|L| \le |H|$, the overall running time is $\mathcal{O}(|G|^4|H|^5)$. \qed

\section{Implementation and Experimental Results}
We have implemented Algorithm \ref{alg-remove-minority-new}. We have used our laptop (2017 MacBook Pro, 2.3 GHz Intel Core i5 processor, compiler version :
Apple LLVM version 10.0.1 (clang-1001.0.46.4)) to test the running time for the following inputs. In these instances, after constructing the corresponding graph $H$, the vertex $0$ of $G$ would have two elements $0,1$ in its list, i.e., $L(0)=\{0,1\}$. 
We wanted to  decide whether there is a homomorphism that maps vertex 0 of $G$ to vertex $1$ of $H$. In all the cases the program shows there is no such  homomorphism. \\
 \\

\vspace{9mm}

\begin{figure}[H]
  \begin{center}
   \includegraphics[scale=0.65]{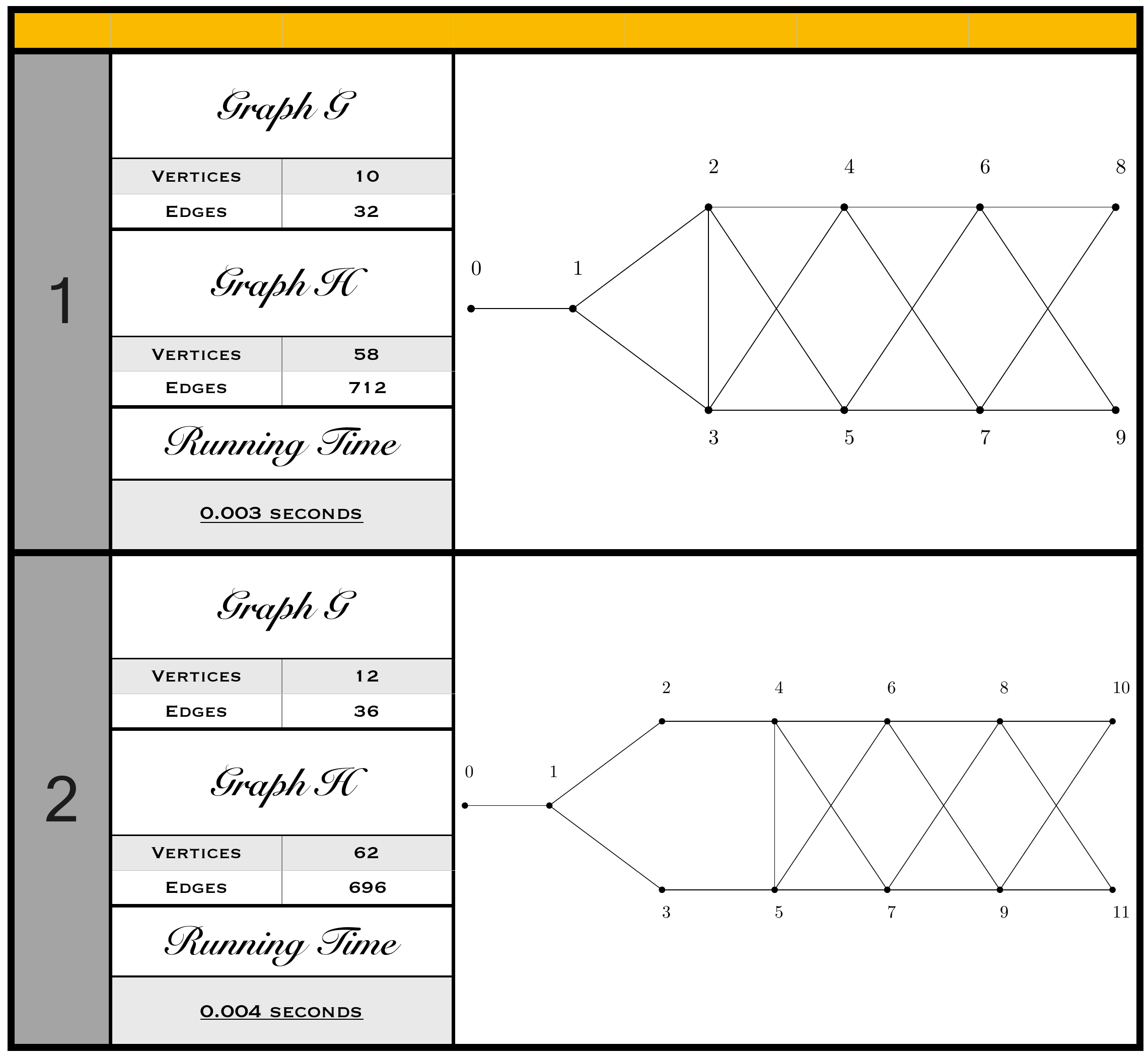}
  \end{center}
  
\label{report-page1}
 \end{figure}

\begin{figure}
  \begin{center}
   \includegraphics[scale=0.6]{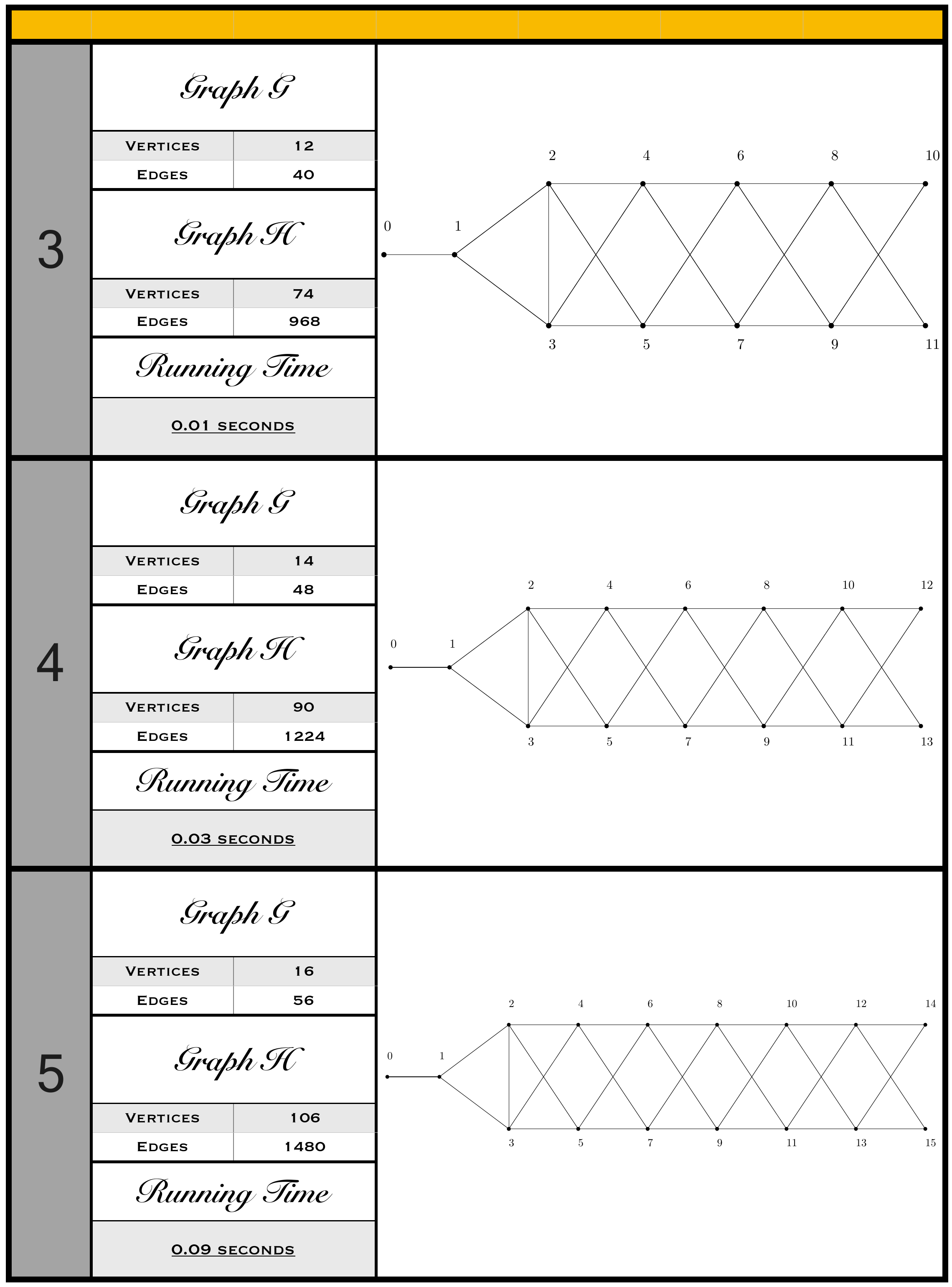}
  \end{center}
  
\label{report-page2}
 \end{figure}

\begin{figure}
  \begin{center}
   \includegraphics[scale=0.6]{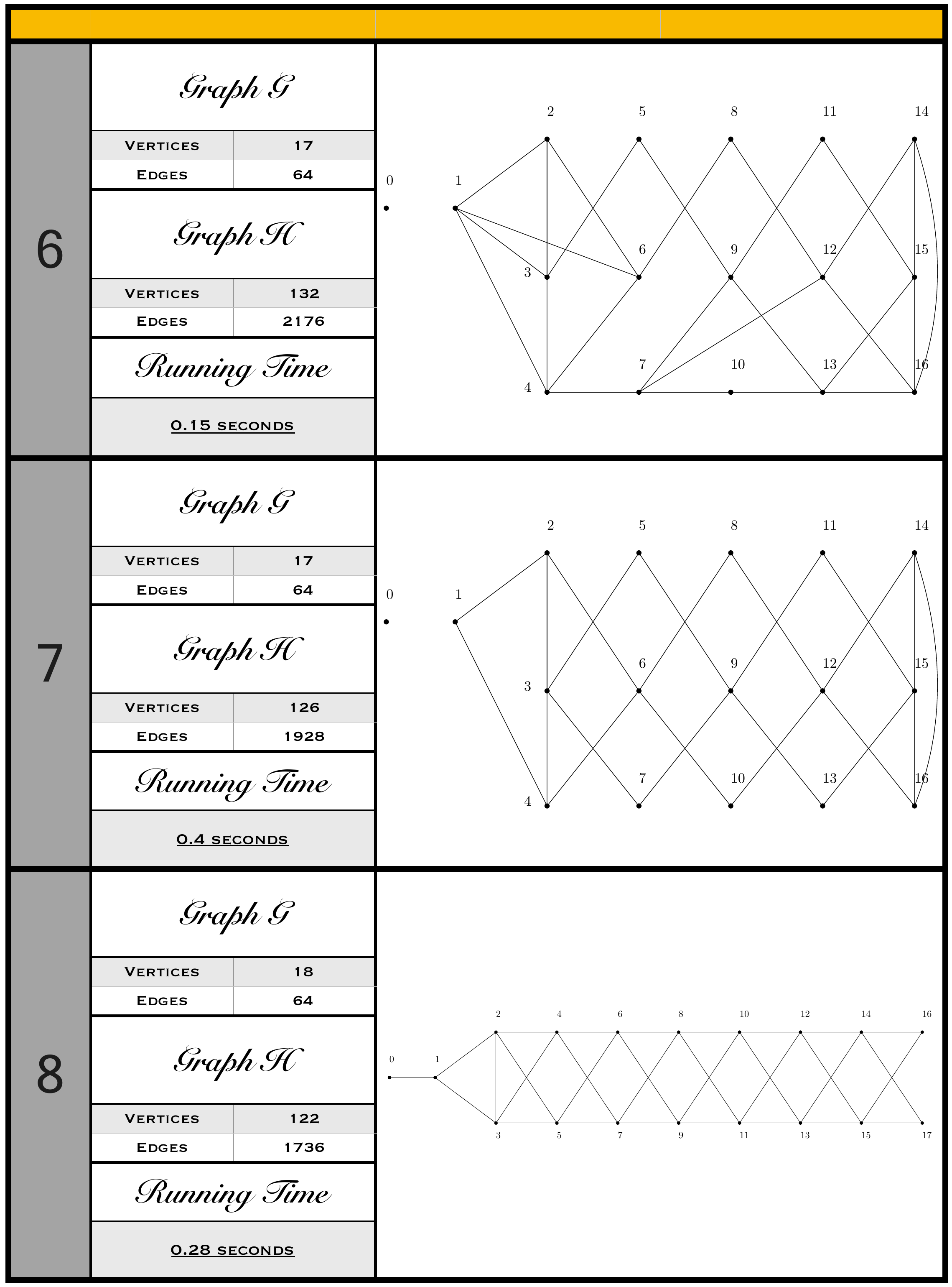}
  \end{center}
 
\label{report-page3}
 \end{figure}

\begin{figure}
  \begin{center}
   \includegraphics[scale=0.6]{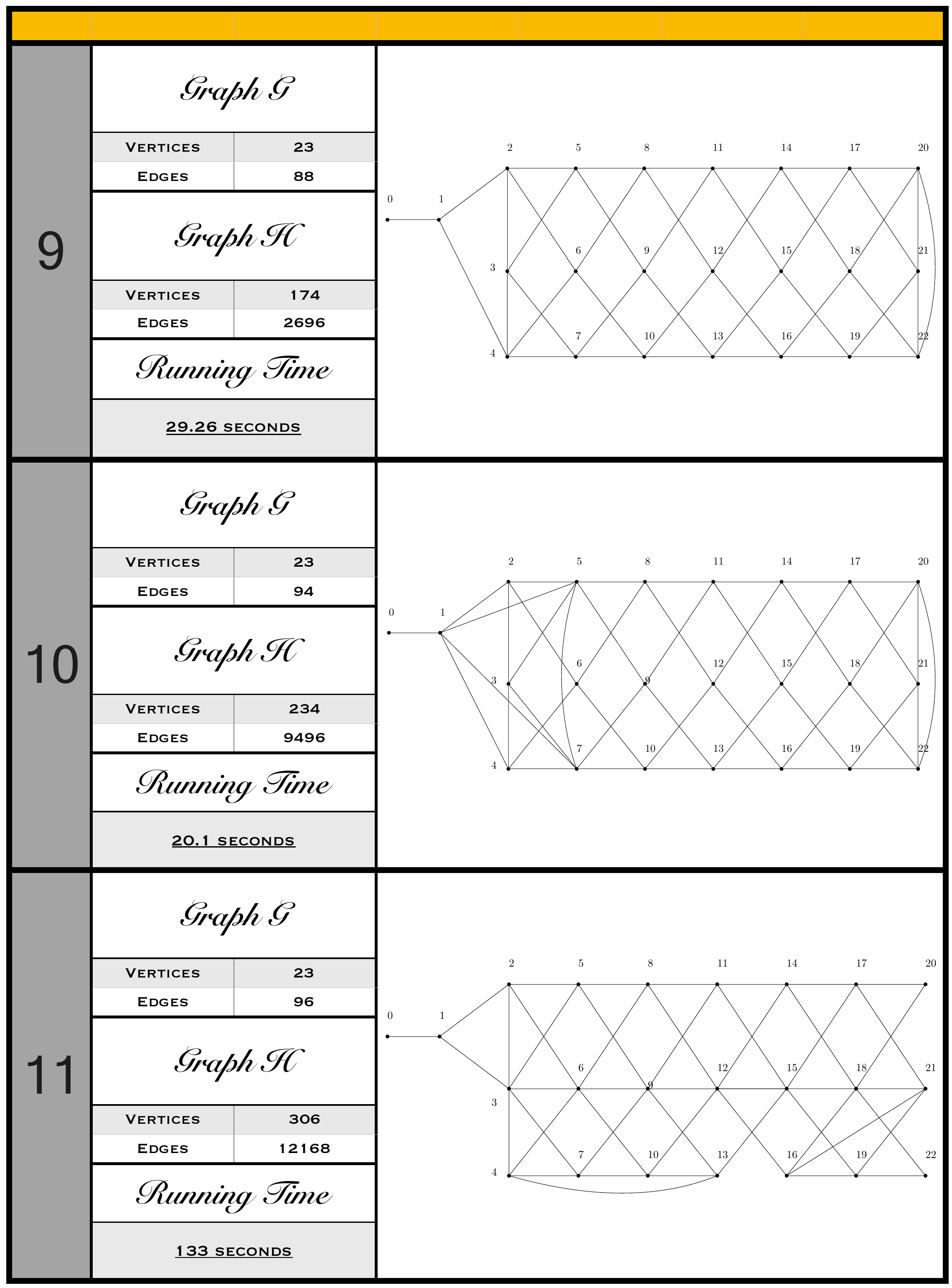}
  \end{center}
 
\label{report-page4}
 \end{figure}

\begin{figure}
  \begin{center}
   \includegraphics[scale=0.6]{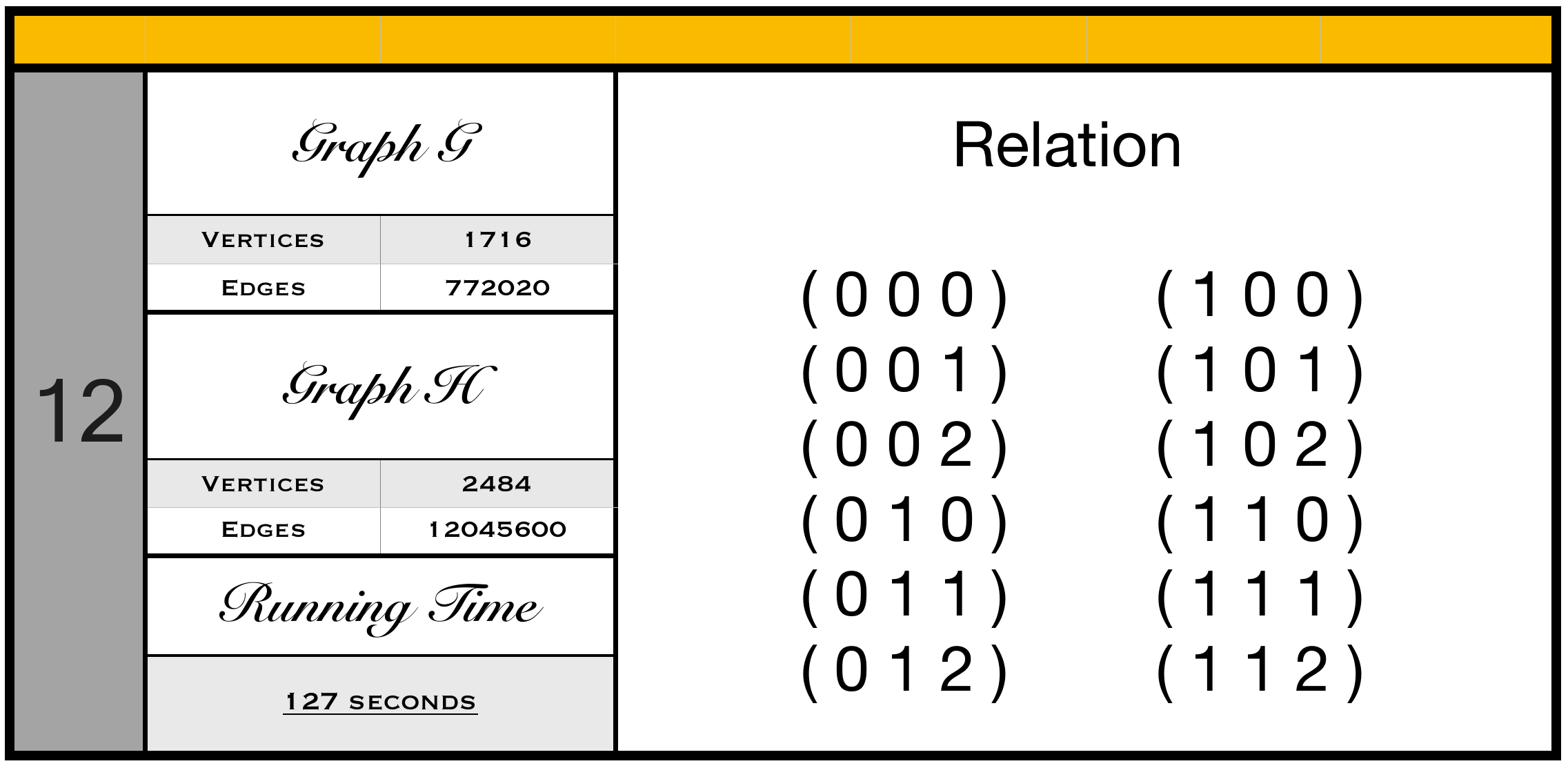}
  \end{center}
 \caption{ Relation and its corresponding graph}
\label{report-page5}
\end{figure}

\newpage
The relation in the Figure \ref{report-page5} admits a minority polymorphism $f(y,x,x)=f(x,y,x)=f(x,x,y)=x$; namely,  $f(x,y,2)=f(x,2,y)=f(2,x,y)=2$ for $x,y$ in $\{0,1\}$. In any other case $f(x,y,z)$ is the minority. 
We construct graph $G$ according to the constructions in section \ref{CSP-to-Graphs}.

\begin{figure}[H]
  \begin{center}
   \includegraphics[scale=0.6]{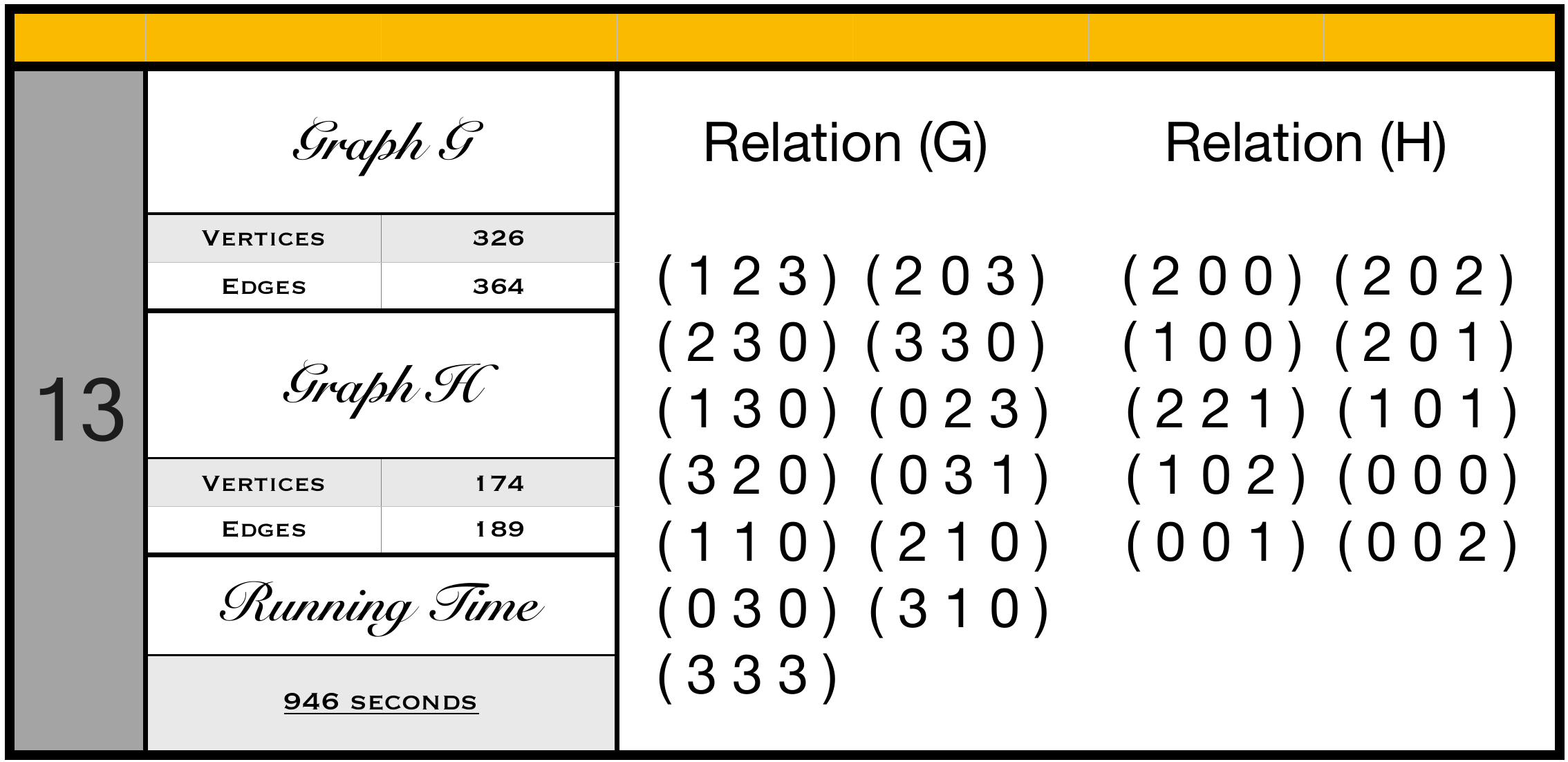}
  \end{center}
 \caption{ Relations $G,H$ and their corresponding digraphs}
\label{report-page6}
\end{figure}

In Figure \ref{report-page6} we construct an instance of the problem according to the construction in \cite{bulin}. 

\section{A Conjecture on Structural Characterization of the Maltsev Polymophisms}

We say a mapping  $h : G \times_L H^3 \rightarrow H$ is a triple consistent Maltsev homomorphism, if for every $xy \in A(G)$, 
and every $a,b,c \in L(x)$, $a',b',c' \in L(y)$ with $(a,b'),(b,b'),(c,c') \in L(x,y)$ the arc $h(x,a,b,c)h(y,a',b',c')$ is in $H$. Moreover, $h(x,a,b,b)=h(x,b,b,a)=a$. In what follows, we are after finding a triple consistent Maltsev assuming that $G \times_L H^3$ admits a Maltsev (list) homomorphism. 

\begin{definition}[Weak Rectangle]
Let $x \in V(G)$, $a \ne b \in L(x)$. We say $(x,a,b)$ induces (is) a weak-rectangle if there exist some $y \in V(G)$, $c \in L(y)$ such that $(a,c),(b,c) \in L(x,y)$. 
\end{definition}

\begin{definition}[Strong Rectangle]
Let $x,y \in V(G)$, and $a \ne b \in L(x)$, $c \ne d \in L(y)$. We say $(x,y,a,b,c,d)$ induces (is) a strong rectangle if $(a,c),(a,d),(b,c),(b,d) \in L(x,y)$. 

\end{definition}

\begin{definition}[Distinguisher]

For every $x \in V(G)$, $a,b \in L(x)$, set $DS(x,a,a,b)=DS(x,b,a,a)=\{b\}$. 

For every $x \in V(G)$, three distinct vertices $a,b,c \in L(x)$, let $DS(x,a,b,c) \subseteq L(x)$ be the set of all vertices $d$ such that :
\begin{itemize}
    \item For every $y \in V(G)$, $\alpha \in L(y)$, if $(a,\alpha),(b,\alpha),(c,\alpha) \in L(x,y)$ then $(d,\alpha) \in L(x,y)$;
    \item For every $y \in V(G)$, $\alpha,\beta \in L(y)$, if $(a,\alpha),(b,\alpha),(c,\beta) \in L(x,y)$ then $(d,\beta) \in L(x,y)$;
    \item For every $y \in V(G)$, $\alpha,\beta \in L(y)$, if $(a,\beta),(b,\alpha),(c,\alpha) \in L(x,y)$ then $(d,\beta) \in L(x,y)$;

\end{itemize}

We call $DS(x,a,b,c)$, the set of distinguishers for $(x,a,b,c)$. 

\end{definition}

\paragraph{Pair consistency on the distinguishers} 
We further prune the distinguishers  as follows. \\
$\forall x,y \in V(G)$, $\forall a,b,c \in L(x)$, $ \forall a',b',c' \in L(y)$ such that $(a,a'),(b,b'),(c,c') \in L(x,y)$ do the following :\\
For every $d \in DS(x,a,b,c)$ if there is no  $d' \in DS(y,a',b',c')$ such that $(d,d') \in L(x,y)$ then remove $d$ from $DS(x,a,b,c)$.   

\begin{conjecture}
Suppose $DS(x,a,b,c) \ne \emptyset$ for every $x \in V(G)$, $a,b,c \in L(x)$ after performing pair consistency on the distinguishers. Then $G \times_L H^3$ admit a Maltsev homomorphism to $H$. Such a homomorphism $h$ can be obtained as follows.

\begin{enumerate}
    \item As long as $\exists \ \ x,a,b,c$ with $DS(x,a,b,a) = \{d\}$ (singleton) then set $h(x,a,b,c)=d$.
    \item As long as there is some $(x,a,b,a)$, $|DS(x,a,b,a)| >1$, set $DS(x,a,b,a)=\{a\}$ and 
    perform pair consistency on the distinguishers.
    \item If there exists some $x,a,b,c$, $a \ne c$, and $|DS(x,a,b,c)| >1$. Then set $h(x,a,b,c)=d$
    for some $d \in DS(x,a,b,c)$ and perform pair consistency on the distinguishers.
\end{enumerate}
\end{conjecture}

\begin{remark}
 We couldn't prove the conjecture. But, we have run a computer program and tested the algorithm in this section on some random graphs as well on the graphs in the previous section. All support the conjecture. 
\end{remark}

\subparagraph*{Acknowledgement :}
We would like to thank  Akbar Rafiey, and Pavol Hell for so many helpful discussions and their useful comments.

\end{document}